\documentclass[pdflatex,sn-mathphys-num]{envelope_optimization}

\usepackage{multirow}%
\usepackage{amsthm}%
\usepackage{mathrsfs}%
\usepackage[title]{appendix}%
\usepackage{textcomp}%
\usepackage{manyfoot}%
\usepackage{booktabs}%
\usepackage{algorithm}%
\usepackage{algorithmicx}%
\usepackage{algpseudocode}%
\usepackage{listings}%

\theoremstyle{thmstyleone}%
\theoremstyle{thmstyletwo}%

\theoremstyle{thmstylethree}%

 \usepackage{layouts}
\usepackage{graphicx, subfigure}
\usepackage{dcolumn}
\usepackage{bm}
\usepackage{mathtools}
\usepackage{amsmath,amssymb,amsfonts}%
\usepackage{physics}

\usepackage{pgfplots}
\usepackage{float}
\usepgfplotslibrary{fillbetween}
\pgfplotsset{compat=1.16}
\usetikzlibrary{fit,backgrounds,positioning,shapes,shapes.callouts}
\usepackage[scaled=0.83]{helvet}
\usepackage{marvosym,pifont}
\usepackage[T1]{fontenc}
\usepackage[utf8]{inputenc}
\usepackage{multirow}
\usepackage{placeins}
\usepackage{chngcntr}

\usepackage{hyperref}

\usepackage{ulem}

\renewcommand{\thefigure}{\arabic{figure}}

\usepackage{xcolor}
\usepackage{colortbl}
\usepackage{array}
\newcolumntype{C}[1]{>{\centering\arraybackslash}p{#1}}

\definecolor{blue}{rgb}{0.0, 0.0, 1.0}
\definecolor{red}{rgb}{1.0, 0.0, 0.0}
\definecolor{lightblue}{rgb}{0.6, 0.6, 1.0}
\definecolor{lightred}{rgb}{1.0, 0.6, 0.6}
\definecolor{white}{rgb}{1.0, 1.0, 1.0}

\newcommand{\colorvalue}[1]{%
	\ifdim #1 pt > 0.8pt
	\cellcolor{red!90} \textcolor{white}{#1} 
	\else \ifdim #1 pt > 0.7pt
	\cellcolor{red!80} \textcolor{white}{#1} 
	\else \ifdim #1 pt > 0.6pt
	\cellcolor{red!70} \textcolor{white}{#1} 
	\else \ifdim #1 pt > 0.5pt 
	\cellcolor{red!60} \textcolor{white}{#1} 
	\else \ifdim #1 pt > 0.4pt
	\cellcolor{red!50} \textcolor{white}{#1} 
	\else \ifdim #1 pt > 0.3pt
	\cellcolor{red!40} \textcolor{white}{#1} 
	\else \ifdim #1 pt > 0.2pt
	\cellcolor{red!30} \textcolor{black}{#1} 
	\else \ifdim #1 pt > 0.1pt
	\cellcolor{red!20} \textcolor{black}{#1} 
	\else \ifdim #1 pt > 0.01pt
	\cellcolor{red!10} \textcolor{black}{#1} 
	\else \ifdim #1 pt > 0.01 pt
	\cellcolor{white} \textcolor{black}{#1}
	\else \ifdim #1 pt > -0.01pt
	\cellcolor{lightblue!10} \textcolor{black}{#1} 
	\else \ifdim #1 pt > -0.1pt
	\cellcolor{lightblue!20} \textcolor{black}{#1} 
	\else \ifdim #1 pt > -0.2pt
	\cellcolor{lightblue!30} \textcolor{black}{#1} 
	\else \ifdim #1 pt > -0.3pt
	\cellcolor{lightblue!40} \textcolor{black}{#1} 
	\else \ifdim #1 pt > -0.4pt
	\cellcolor{lightblue!50} \textcolor{black}{#1} 
	\else \ifdim #1 pt > -0.5pt
	\cellcolor{lightblue!60} \textcolor{black}{#1} 
	\else \ifdim #1 pt > -0.6pt
	\cellcolor{lightblue!70} \textcolor{black}{#1} 
	\else \ifdim #1 pt > -0.7pt
	\cellcolor{lightblue!80} \textcolor{black}{#1} 
	\else \ifdim #1 pt > -0.8pt
	\cellcolor{lightblue!90} \textcolor{black}{#1} 
	\else
	\cellcolor{blue!100} \textcolor{white}{#1} 
	\fi \fi \fi \fi \fi \fi \fi \fi \fi \fi \fi \fi \fi \fi  \fi \fi \fi \fi \fi
}

\begin{document}

\title[Article Title]{Pareto-optimality of pulses for robust population transfer in a ladder-type qutrit}


\author[1]{\fnm{John J.} \sur{McCord}}

\author*[1]{\fnm{Marko} \sur{Kuzmanovi\'c}}\email{marko.kuzmanovic@aalto.fi}

\author[1]{\fnm{Gheorghe Sorin} \sur{Paraoanu}}

\affil*[1]{\orgdiv{QTF Centre of Excellence, Department of Applied Physics}, \orgname{Aalto University}, \city{Aalto}, \postcode{FI-00076}, \country{Finland}}




\abstract{	Frequency-modulation schemes offer an alternative to standard Rabi pulses for realizing robust quantum operations. In this work, we investigate short-duration population transfer between the ground and first excited states of a ladder-type qutrit, with the goal of minimizing leakage into the second excited state. Our multiobjective approach seeks to reduce the maximum transient second-state population and maximize detuning robustness.  Inspired by two-state models—such as the Allen-Eberly and Hioe-Carroll models—we extend these concepts to our system, exploring a range of pulse families, including those with super-Gaussian envelopes and polynomial detuning functions. We identify Pareto fronts for pulse models constructed from one of two envelope functions paired with one of four detuning functions. We then analyze how each Pareto-optimal pulse parameter influences the two Pareto objectives as well as amplitude robustness.
		}




\maketitle

\section{Introduction}\label{sec1}

Robust population transfer remains a key area of interest in quantum control. Various techniques employing adiabatic passage have been the primary means to this end, engineered to deliver high fidelity and efficiency with minimal dissipation. The principle behind these methods lies in the concept of adiabaticity, which ensures that a system remains in its instantaneous eigenstate throughout its evolution. For example, in  a two-level system with Hamiltonian $\hat{H} = \frac{\hbar}{2}(\Omega \hat{\sigma}_x + \Delta \hat{\sigma}_z)$, the energy eigenvalues are $E_{\pm} = \pm \frac{\hbar}{2}\sqrt{\Delta^2 + \Omega^2}$, corresponding to eigenstates
$\vert - \rangle =  \cos(\theta/2) \vert  0\rangle - \sin(\theta/2) \vert  1\rangle$ and 
$\vert + \rangle = \sin(\theta/2) \vert  0\rangle + \cos(\theta/2) \vert  1\rangle$, where $\theta$ is the so-called mixing angle and relates to the amplitude and detuning via $\tan (\theta) = \Omega / \Delta$. The system evolves adiabatically if $\dot{\theta}/\Delta E  \ll 1$, which is satisfied when $\frac{\norm{\dot{\Omega}\Delta - \Omega \dot{\Delta}}}{(\Omega^{2} + \Delta^{2})^{3/2}} \ll 1$ \cite{Vitanov_2001}. When this condition is met, population transfer occurs through gradual evolution along eigenstates, avoiding unwanted transitions.

	Rapid adiabatic passage (RAP) is the staple technique for robust population inversion. RAP was first realized in NMR, where the spins on the magnetic moments are inverted by sweeping the drive field frequency across the resonance \cite{Bloch1946}. It is \textit{rapid} in the sense that the frequency sweep is faster than the relaxation time of the spins, or in the case of circuit QED, the relaxation time of the qubit.  RAP is very versatile in that it can be implemented on any two-level quantum system with an external drive field. As such, it has found many applications in atomic, molecular, and optical physics \cite{Loy1974, Grischkowsky1976, Caussyn1994, Noel2012}, as well as in chemical reactions \cite{Kral2007}, interferometry \cite{Weitz1994, Kovachy2012}, and circuit QED \cite{Wei2008, Nie2010}. In circuit QED, the cited works utilize Stark-chirped rapid adiabatic passage (SCRAP), a variety of RAP, in which a slowly varying Stark shift sweeps the energy of the target state through resonance to induce complete population transfer \cite{Yatsenko1999}.

	Beyond these implementations, RAP has also inspired a number of fast and robust quantum control protocols. For example, the Roland–Cerf protocol introduced a locally adiabatic schedule that slows near the avoided crossing to maintain high fidelity while reducing overall evolution time \cite{Roland_2002}. This idea has since been experimentally realized and extended in various contexts, including high-fidelity quantum driving \cite{Bason_2012}, as well as resonant shortcuts to adiabatic passage using only longitudinal ($z$-field) control \cite{Stefanatos_2019}. Similarly, RAP-inspired strategies have been employed to implement qubit gates using longitudinal ($\sigma_z$) control alone \cite{Martinis_2014}.
	
	In the context of RAP, pulse shaping is essential for ensuring efficient and robust population transfer. Gaussian pulses are often employed \cite{Demirplak_2003, Rangelov_2012}, in part because they have been shown to reduce leakage beyond the computational space \cite{Steffen_2003}. Super-Gaussian pulses have also been utilized to enhance control over transition dynamics. For instance, their flatter tops not only suppress spectral leakage and limit off-resonant excitations but also enhance detuning robustness, particularly near the pulse center—an effect especially beneficial for adiabatic or RAP-like protocols \cite{Kuzmanovic_2024}. 	
	
	A central goal in quantum control of weakly anharmonic systems, such as the transmon, is to suppress leakage outside the computational subspace—that is, to confine the population dynamics to the lowest two energy levels. When driving the ground-to-first-excited-state transition $\mathrm{g-e}$ with a resonant microwave pulse, the close proximity of the $\mathrm{e-f}$ transition frequency means that higher levels cannot be ignored. The spectral width of the pulse—particularly if it is short and strong—can unintentionally drive the $ \mathrm{e-f}$ transition, leading to population leakage out of the computational subspace.
	To mitigate this, techniques such as derivative removal by adiabatic gate (DRAG) pulses are used, which shape the drive to cancel unwanted transitions \cite{Motzoi_2009, Hyyppa_2024}. Alternatively, using longer pulses reduces spectral overlap but can increase sensitivity to decoherence \cite{Khodjasteh_2011}. Ultimately, achieving high-fidelity control in such systems requires carefully balancing gate speed, spectral selectivity, and pulse shaping to minimize leakage while maintaining fast, accurate operations \cite{Werninghaus_2021, Babu_2021}.

	In this work, we pursue this goal by building on rapid adiabatic passage (RAP) methods and our previous work involving high-fidelity qubit control using phase-modulated pulses \cite{Kuzmanovic_2024}. Since the phase modulation we employ is equivalent to a frequency modulation through an appropriate choice of frame rotation, we hereafter refer to the scheme as one that utilizes frequency modulation.  In the simplest case, this drive frequency is modulated by adding to the qubit frequency an instantaneous linear detuning $\Delta(t) = 2\Delta_{\rm{max}} t/T$, parametrized in terms of modulation depth $\Delta_{\rm{max}}$ and pulse duration $T$, with $t \in [-T/2, T/2]$. 
	Furthermore, following our frequency-modulation technique, we employ a single continuously modulated pulse and explore different RAP-like models on a three-level ladder system with states $\{|\rm{g}\rangle, |\rm{e}\rangle, |\rm{f}\rangle \}$, and seek to minimize leakage into the second excited state $|\rm{f}\rangle$ when driving population from the ground state to the first excited state. Specifically, we wish to minimize the maximum transient second excited state population $\max_{t}(p_{\rm{f}})$, with $t \in [-T/2, T/2]$, and simultaneously maximize the detuning robustness at a threshold of the final first excited state population $p_{\rm{e}}(T/2) \geq 0.99$.  As this is a multiobjective optimization problem, we seek to find the so-called Pareto front which meets our criteria. A Pareto front is a solution to the multiobjective optimization in which it proves impossible to improve one objective without making the other worse. Pareto fronts have found applications in a plethora of disciplines including economics \cite{Newberry1984}, traffic regulation \cite{Jiao2016}, machine learning \cite{Jin2008}, and bioinformatics \cite{Gonzalez-Sanchez2023}. In physics, Pareto's methodology has been employed to find the extrema of observables as a form of optimal quantum control \cite{Chakrabarti2008, Sun2017}. 
	
		This paper is organized as follows: In Sect.~\ref{sec:Meth} we describe the methodology for our frequency-modulated robust population transfer scheme, introduce the pulse models used in this work, and detail the underlying methodology for the Pareto optimization. In Sect.~\ref{sec:Res}, we present the Pareto fronts for the different models, each at two pulse durations, and assign an effective limit to the fronts by utilizing the Landau-Zener formula. We perform parameter analysis in Sect.~\ref{sec:Anal}, and ascertain the most impactful parameters for each of our objective outcomes. Finally, we summarize the key findings of this work in Sect.~\ref{sec:Con}.

	\section{Methodology}
	\label{sec:Meth}

	This section has three objectives. First, we describe the frequency modulation scheme adopted in this work. Second, we present the pulse models explored, along with the parameters targeted for optimization. Finally, we describe the algorithm employed to compute the Pareto front.
	
	\subsection{The frequency-modulation scheme adopted to a ladder qutrit}
	
\begin{figure}[h]
		\centering
		\includegraphics[width=0.7\linewidth]{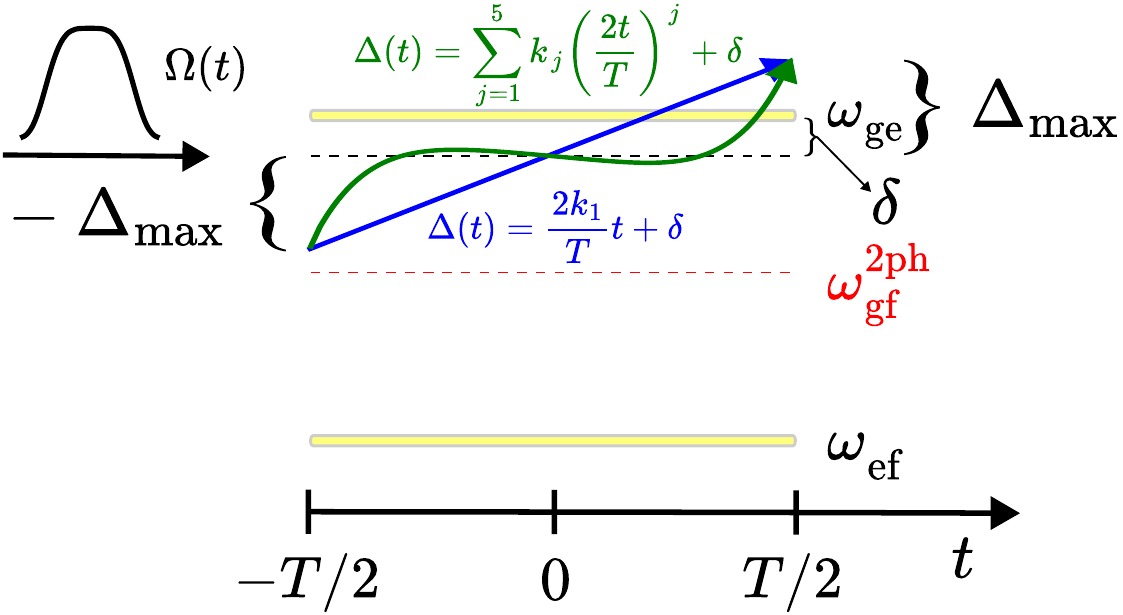}
		\caption{The energy diagram of our frequency-modulation scheme shown for two different detuning functions, one linear (blue) and one quintic (green), for time $t \in [-T/2, T/2]$. The transition frequencies are $\omega_{\rm{ge}}$ and $\omega_{\rm{ef}}$; the modulation depth is denoted by $\Delta_{\rm max}$; and the frequency offset is $\delta$. The two-photon transition frequency $\omega_{\rm{gf}}^{\rm 2ph}$ is exactly halfway between $\omega_{\rm{ge}}$ and $\omega_{\rm{ef}}$.} %
		\label{fig-energy}
	\end{figure}

	We employ a single, continuously modulated drive pulse with Rabi amplitude $\Omega (t)$ and detuning $\Delta (t)$ with respect to the first transition in order to emulate rapid adiabatic passage between the ground state $|\rm{g}\rangle$ and the first excited state $|\rm{e}\rangle$. An energy diagram of the modulation scheme is shown in Fig.~\ref{fig-energy}. The diagonal part of the Hamiltonian in Eq.~(\ref{eq:Ham}) corresponds to a ladder-type qutrit with energy level separations $\omega_{\rm{ge}}$ and $\omega_{\rm{ef}}$ in the frame co-rotating with the drive frequency. For a transmon qutrit, the ratio of the Josephson energy $E_{J}$ to the charging energy $E_{C}$ is very large, and in the extreme limit $E_{J}/E_{C} \rightarrow \infty$, the transmon energy levels approach those of a harmonic oscillator for low-energy states, with nearly constant but slightly decreasing spacing due to residual anharmonicity. The qutrit is driven by a microwave field with frequency $\omega_{d}(t)$.

	We consider a rotating frame such that for the ground state $\vert \rm{g} \rangle$ the energy is shifted by $-\Delta (t)$ and the zero of the energy is set at the first excited state $\vert \rm{e} \rangle$, where $\Delta(t) =\omega_{\rm ge}-\omega_{d}-(d\omega_{d}/dt)t$. The charging energy $E_{C}$ gives rise to the anharmonicity in the energy levels in superconducting qutrits such as the transmon and relates to the transition frequencies via $E_{\rm C} = \hbar \omega_{\rm{ge}} - \hbar \omega_{\rm{ef}}$. Thus, the detuning of the $|\mathrm{f}\rangle$ level is $\Delta (t) - E_{\rm C}/\hbar$. In this work, we set $E_{C} = 0.3$ GHz, which is within the typical range for a transmon \cite{Koch_2007}.
	The ratio of the Rabi frequencies of the two transitions is given by the ratio of the transition electric dipole moments. The dipole operator is proportional to the number operator $\hat{n} = -i \partial /\partial \varphi$ and thus the dipole matrix elements are $g_{n,n+m} \sim \langle n \vert \hat{n} \vert  n + m \rangle $ and consequently, $g_{12}/g_{01} = \sqrt{2}$ and $g_{02} = 0$. 
	
	Combining these considerations, we arrive at the Hamiltonian in the rotating wave approximation, as detailed in Ref.~\cite{Bjorkman_2025}:
	
	\begin{equation}
		\label{eq:Ham}
		H(t) = 
		\hbar\begin{pmatrix}-\Delta(t)&\Omega (t)/2&0 \\
			\Omega (t)/2&0&\Omega (t)/\sqrt{2}\\
			0&\Omega (t)/\sqrt{2}& \Delta (t) - E_{C}/\hbar\end{pmatrix}.
	\end{equation}
	
	\subsection{Pulse models}
	
	The shape of the drive pulse plays a central role in achieving efficient and robust state transfer. Several envelope functions have been explored in two-state and three-state quantum models, each with distinct advantages and limitations \cite{Rosen1932, Demkov1969, Bambini1981, Allen1987, Carroll1986, Morris_1964, Rangelov_2005, Rangelov_2012, Yuan_2023}. Constant-amplitude pulses can drive transitions via Rabi oscillations but require precise timing and are highly sensitive to detuning errors \cite{Rabi1937}. Landau-Zener transitions, which use a linearly time-dependent detuning with a constant coupling, offer analytical solutions but can suffer from unwanted population leakage \cite{Landau1932, Zener1932}. Further, exponential pulses have been applied in time-dependent quantum transition models but often suffer from long tails, reducing efficiency by causing non-negligible residual population transfer outside the intended time window \cite{Demkov1964, Nikitin1970}.
	
	Smoothly varying pulses, such as Gaussian and hyperbolic secant envelopes, provide greater control over transition probabilities. Gaussian pulses minimize spectral side lobes and have been widely used in adiabatic passage schemes \cite{Demirplak_2003, Vasilev2004, Vasilev_2005, Guerin_2011, Rangelov_2012, Yuan_2023}. However, achieving complete population transfer with Gaussian pulses requires precise detuning control. On the other hand, hyperbolic secant pulses are analytically solvable and naturally suited for robust adiabatic and nonadiabatic population transfer, making them widely used in RAP and related techniques \cite{Rosen1932, Demkov1969, Bambini1981, Allen1987, Carroll1986, Ishkhanyan_2014, Ishkhanyan_2015}.
	
	In this work, we focus on two pulse envelopes: generalized super-Gaussian and hyperbolic secant varieties. For appropriately chosen parameters, both envelope types are flat at the top, meaning the pulse amplitude remains approximately constant near the center of the drive, around $t=0$, where the avoided crossing typically occurs. This flat-top structure grants a greater degree of detuning robustness by maintaining a high Rabi frequency over the critical transition region, thereby extending the adiabatic window and improving population transfer even under moderate detuning errors \cite{Kuzmanovic_2024}. The super-Gaussian envelope allows for greater control over the transition window by flattening the peak and steepening the decay, which can simultaneously be advantageous in terms of requiring lower amplitudes. Meanwhile, the hyperbolic secant envelope serves as a benchmark due to its well-established analytical properties. These envelope functions are defined as follows:
	
	\begin{equation} 
		\Omega(t) = \Omega_{0}\Bigg{[}\frac{e^{\beta|2t/T|^n} - e^{\beta}}{1 - e^\beta}\Bigg{]}\;,
	\end{equation}
	\newline
	for the super-Gaussian envelope and
	
	\begin{equation} \Omega(t) = \Omega_{0}\Bigg{[}\frac{\sech(\epsilon|\frac{2t}{T}|^n) - \sech(\epsilon)}{1-\sech(\epsilon)}\Bigg{]}\;,
	\end{equation}
	\newline
	for the hyperbolic secant envelope, where in both cases we consider times $ t \in [-T/2, T/2]$. We have specifically tailored our envelope functions such that, for the super-Gaussian (hyperbolic secant) envelope, the subtraction by $e^{\beta}$ ($\sech(\epsilon)$) and division by $1 - e^{\beta}$ ($1 - \sech(\epsilon)$) ensure that $\Omega(0) = \Omega_{0}$ and $\Omega(\pm T/2) = 0$. Without the aforementioned normalization, we would have $\Omega(\pm T/2) = \Omega_{0} e^{\beta}$ for the super-Gaussian envelopes and $\Omega(\pm T/2) = \Omega_{0} \sech(\epsilon)$ for the hyperbolic secant envelopes.

	By employing these two pulse shapes, we aim to explore their effects on state transfer efficiency and robustness, particularly in the context of our frequency-modulation scheme adopted to a three-level system. Each of these envelopes contain three parameters which we seek to multiobjectively optimize with respect to the minimization of the maximum second excited state population $\max_{t}(p_{\rm{f}})$ reached across the pulse duration $ t \in [-T/2, T/2]$ and (simultaneously) with respect to the detuning robustness defined as the frequency width of the first excited state population at the threshold $p_{\rm{e}}(T/2) \geq 0.99$ with a local maximum of $p_{\rm{e}}(T/2) \geq 0.999$ within this width. We emphasize transient rather than end-of-pulse leakage because in a transmon, higher-order corrections in the weakly anharmonic spectrum cause temporary population excursions into non-computational states. Even if these populations relax back by the end of the pulse, such excursions still degrade gate performance and may couple to uncontrolled decay and decoherence channels. Moreover, the condition $p_{\rm{e}}(T/2) \geq 0.999$ ensures that the final population of the $\ket{\mathrm{f}}$ state is on the order of $10^{-3}$ or less, an order of  magnitude smaller than the typical observed transient population.  The robustness threshold values are chosen as a balance between experimental relevance and numerical stability, whereas a stricter requirement (e.g., $0.9999$) would amplify sensitivity to fluctuations without qualitatively changing the optimization landscape.

	The envelope parameters that we seek to multi-objectively optimize are the amplitude $\Omega_{0}$, envelope order $n$, and the parameter $\beta$ (for super-Gaussian pulses) or $\epsilon$ (for hyperbolic secant pulses), which determine the envelope width and steepness. The envelope order $n$ generalizes the super-Gaussian envelopes in that if $n = 2$ we recover a Gaussian profile, $n=4$ corresponds to the so-called super-Gaussian variety, and as $n \rightarrow \infty$ the pulse approaches a rectangular form. Amplitudes $\Omega_{0}$ are searched for in the range $[0, 0.1]$ GHz, consistent with experimental constraints that help mitigate parasitic effects such as leakage into $\vert \rm{f}\rangle$ and higher states, as well as AC Stark shifts, which are known to degrade qubit coherence and fidelity in similar systems \cite{Shillito_2022}. The envelope order $n$ is scanned over the range $(0, 10]$, while $\beta$ and $\epsilon$ are searched within $[-6, 0)$ and $(0, 6]$, respectively.

	Additionally, we consider four detuning models and assign each to the two envelope types. Our first detuning model is linear and the model name, when assigned to the super-Gaussian envelope, is SG1 and SECH1 when assigned to the hyperbolic secant envelope. For subsequent models, we hereafter adopt the prefixes SG and SECH for the model names. 
		
	The first detuning model has the form 
	\begin{equation}
		\Delta(t) = \frac{2k_{1}}{T}t + \delta\;,
		\label{eq:linear}
	\end{equation}
	where the prefactor $k_{1}$ determines the slope of our detuning function and is a parameter we optimize by searching in the range $k_{1} \in [-0.3, 0.3]$ GHz. Since Eq.~(\ref{eq:linear}) is a linear detuning function, the parameter $k_{1}$ is the modulation depth, which typically does not exceed half the charging energy, i.e., $E_{C}/2 = 0.15$ GHz in this work. Further, we center our modulation about the frequency offset $\delta$, i.e., by sweeping $(2\Delta_{\rm max}/T) t$ from $t = -T/2$ to $t = T/2$. 
	
	We also explore a quintic detuning function, 
	
	\begin{equation}
		\Delta(t) =	\sum \limits_{j = 1}^{5}k_{j}(2t/T)^j + \delta,
		\label{eq:quintic}
	\end{equation}
	where $k_{j > 1}$ are each searched for in the range $[-0.1, 0.1]$ GHz. For this detuning function, the effect of the parameter $k_{1}$ is the same as in Eq.~(\ref{eq:linear}); whereas, the prefactor for the quadratic term, $k_{2}$, makes the detuning function convex (upward-bending) when $k_{2} > 0$ and concave (downward-bending) when $k_{2} < 0$. Further, the parameter $k_{3}$, the cubic prefactor, introduces asymmetry and has the effect of making the right side ($t > 0$) of the detuning function steeper and the left side ($t < 0$) shallower when $k_{3} > 0$ and vice versa when $k_{3} < 0$. Still further, the quartic ($k_{4}$) and quintic ($k_{5}$) prefactors have similar effects to $k_{2}$ and $k_{3}$, respectively, but with stronger effects at larger $\abs{t}$. Overall, the symmetry properties are governed by the parity of each term: the even-powered terms ($k_{2}$, $k_{4}$) contribute symmetric components about $t = 0$, while the odd-powered terms ($k_{1}$, $k_{3}$, $k_{5}$) contribute antisymmetric components. A detuning function composed purely of odd-powered terms is antisymmetric about the origin, but adding even-powered terms breaks this antisymmetry, resulting in an overall asymmetric function. Thus, the inclusion of both even and odd terms enables finer control over both the shape and symmetry of the detuning profile.
	
	We also examine detuning functions that incorporate the hyperbolic tangent. Motivated by the Allen–Eberly \cite{Allen1987} and Hioe–Carroll \cite{Carroll1986} models—both of which employ hyperbolic secant pulse envelopes—we respectively consider the following forms: 
	\begin{equation}
		\Delta (t) = k_{1} \tanh(2\gamma_{1} t/T) + \delta
		\label{eq:ae}
	\end{equation}
	and 
	
	\begin{equation}
		\Delta(t) = k_{1}\tanh(2\gamma_{1} t/T) + k_{2}(\sech(2\gamma_{2} t/T) - 1) +  \delta,
		\label{eq:hc}
	\end{equation}
	\linebreak
	where we subtract by $k_{2}$ in Eq.~(\ref{eq:hc}) to ensure that we satisfy the condition $\Delta(0) - \delta = 0$, which is naturally satisfied for the other detuning functions we consider. For these detuning functions, the parameters $k_{1}$ and $k_{2}$ are searched for in the aforementioned ranges $k_{1} \in [-0.3,0.3]$ GHz and $k_{2} \in [-0.1, 0.1]$ GHz; whereas, the parameters $\gamma_{1}$ and $\gamma_{2}$ are each searched for in the range $(0, 8]$. The parameter $k_{1}$ acts as a scaling factor which symmetrically affects the value at which the detuning saturates, i.e., the detuning asymptotically approaches $\pm k_{1}$ since $\tanh(x) \rightarrow \pm 1$ when $\abs{x}$ is large. However, 
	$\gamma_{1}$ controls how quickly the detuning transitions from negative to positive values, i.e., the steepness of the slope. Small values of $\gamma_{1}$ yield a near-linear detuning and large values yield a sharper transition near $t = 0$, with extremely large values producing a step-like detuning. Since Eq.~(\ref{eq:hc}) is just Eq.~(\ref{eq:ae}) plus an additional term, the effect of these parameters is the same in both detuning functions. Moreover, the additional term in Eq.~(\ref{eq:hc}) introduces a dip in the detuning function around $t = 0$ and increasing $k_{2}$ makes the dip more pronounced. The parameter $\gamma_{2}$ affects the localization of the dip, with a larger $\gamma_{2}$ yielding a dip more concentrated around $t = 0$ and a smaller $\gamma_{2}$ making it spread out over a longer time.    
	
	As we have previously introduced the shorthand notations SG and SECH for our two envelope types, we further systematically distinguish our models by referring to Eqs.~(\ref{eq:linear}), (\ref{eq:quintic}), (\ref{eq:ae}), and (\ref{eq:hc}) as detunings of type 1, type 2, type 3, and type 4, respectively. We then create our model names by agglutinating the SG or SECH suffix to a particular detuning model as shown in Table~\ref{tab-models}. Thus, in this work, we consider a total of eight models: four with super-Gaussian envelopes and four with hyperbolic secant envelopes. These models each satisfy the conditions $\Omega(0) = \Omega_{0}$, $\Omega(\pm T/2) = 0$, $\Delta(0) = \delta$ and are evaluated at pulse durations of $T = 50$ ns and $T = 200$ ns. These two pulse durations span a range of experimentally relevant gate times in superconducting qubit hardware. Shorter durations are motivated by the need for fast gate operations in near-term devices, whereas longer durations allow greater detuning robustness and reduced leakage into higher excited states. The Pareto-optimal smooth-envelope pulses considered in this work exhibit spectral bandwidths dominated by frequency modulation, instead of the pulse duration itself. Their half-power bandwidth is typically on the order of $\pm 100$ MHz, well within the capabilities of contemporary RF signal generators. As an example, in transmon-based systems, we have previously shown that sufficiently long pulses driven close to resonance with modest modulation depths can exhibit detuning robustness with a negligible final second excited state population, thereby isolating the qubit subspace despite the transmon’s weak anharmonicity \cite{Kuzmanovic_2024}.
	
	\begin{table}[h!]
    \centering
    \scalebox{1.15}{ 
    \renewcommand{\arraystretch}{1.5} 
    \begin{tabular}{|c|c|c|}
        \hline
        \multicolumn{3}{|c|}{Two-state-inspired models} \\
        \hline
        Model & Envelope $[\Omega(t)/\Omega_{0}]$ & Detuning $[\Delta(t) - \delta]$\\
        \hline 
        &&\\
        SG1 & $\left( \frac{e^{\beta|2t/T|^n} - e^{\beta}}{1 - e^\beta}\right)$ & $2k_{1}t/T$\\
        &&\\
        SG2 & $\left( \frac{e^{\beta|2t/T|^n} - e^{\beta}}{1 - e^\beta}\right)$ & $\sum \limits_{j = 1}^{5}k_{j}(2t/T)^j$\\
        &&\\
        SG3 & $\left( \frac{e^{\beta|2t/T|^n} - e^{\beta}}{1 - e^\beta}\right)$ & $k_{1}\tanh(2\gamma_{1} t/T)$\\
        &&\\
        SG4 & $\left( \frac{e^{\beta|2t/T|^n} - e^{\beta}}{1 - e^\beta}\right)$ & $k_{1}\tanh(2\gamma_{1} t/T)$\\ &&  $ +  k_{2}(\sech(2\gamma_{2} t/T) - 1) $\\
        &&\\
        SECH1 & $\left(\frac{\sech(\epsilon|2t/T|^n) - \sech(\epsilon)}{1-\sech(\epsilon)}\right)$ &  $2k_{1}t/T$\\
        &&\\
        SECH2 & $\left(\frac{\sech(\epsilon|2t/T|^n) - \sech(\epsilon)}{1-\sech(\epsilon)}\right)$ & $\sum \limits_{j = 1}^{5}k_{j}(2t/T)^j$\\
        &&\\
        SECH3  & $\left(\frac{\sech(\epsilon|2t/T|^n) - \sech(\epsilon)}{1-\sech(\epsilon)}\right)$ & $k_{1}\tanh(2\gamma_{1} t/T)$\\
        &&\\
        SECH4  &$\left(\frac{\sech(\epsilon|2t/T|^n) - \sech(\epsilon)}{1-\sech(\epsilon)}\right)$ & $k_{1}\tanh(2\gamma_{1} t/T)$\\ &&  $ +  k_{2}(\sech(2\gamma_{2} t/T) - 1) $\\
        \hline
    \end{tabular}
    }
   \newline
    \newline
    \caption{The two-state-inspired models used in this work, each consisting of a pulse envelope function $\Omega(t)$ and detuning function $\Delta(t)$.  Each envelope function has a total of three parameters, i.e., $[ \Omega_{0}, n, \beta\;(\epsilon)]$ for SG (SECH) envelopes. Here, the detuning prefactors are denoted as $k_1$, $\cdots$, $k_5$, with detuning models of type 1 and 3 only using $k_{1}$, type 4 using $k_{1}$ and $k_{2}$, and type 2 using all 5. The parameters $\gamma_1$ and $\gamma_2$ only appear in the detuning functions of types 3 and 4. Each model satisfies the conditions $\Omega(0) = \Omega_{0}$, $\Omega(\pm T/2) = 0$, and $\Delta(0) = \delta$.}
    \label{tab-models}
\end{table}
	
	In the following subsection, we discuss the Pareto-optimization technique for finding parameters that, when inserted into the matrix elements of Eq.~(\ref{eq:Ham}), guide the system to adiabatically follow its lowest eigenstate, thus facilitating robust population transfer from $\vert \rm{g}\rangle$ to $\vert \rm{e}\rangle$.
	
	\subsection{Pareto-optimization}
	
	The conundrum of determining the optimal pulse parameters for our models arises from the fact that a pulse that maximizes detuning robustness $\Delta_{\rm{rob}}$ at a given threshold of $p_{\rm{e}}(T/2)$ will generally not minimize the maximum transient second-excited state population, $\max_{t}(p_{\rm{f}})$, and vice versa. This creates a multiobjective optimization (MOO) problem, where we must balance these competing objectives. To address this trade-off, we apply evolutionary computation methods to identify the best solutions. Techniques such as genetic algorithms and estimation of distribution algorithms have been successfully used in complex optimization problems across various fields, including evacuation planning \cite{Garrett2006} and robotic control optimization \cite{Liang2024}.
	Genetic algorithms operate by mimicking the principles of natural selection to iteratively evolve a population of candidate solutions toward optimality. The process begins with the generation of an initial population that satisfies predefined bounds and constraints. Each individual in the population is evaluated using an objective function to assign a fitness score. The algorithm then progresses through iterative stages, where parents for the next generation are selected based on their fitness \cite{Miller1995}.

	In solving this MOO, we rely on the Pareto front, which represents the optimal trade-off between objectives in the objective space. Pareto fronts have been used in multiobservable quantum control, where it was shown that the smoothness and convexity of the Pareto front depend on the commutativity of the observables \cite{Sun2017}. To find the Pareto front numerically, we use MATLAB's \texttt{gamultiobj()} function, which implements a controlled, elitist genetic algorithm based on the non-dominated sorting genetic algorithm (NSGA-II) \cite{Deb2002}.

	Building on these general principles, NSGA-II enhances the process with additional layers of sophistication, such as non-dominated sorting and crowding distance measures. After scoring the offspring based on their objective function values and feasibility, the current population and offspring are merged into a single set. This set is sorted into non-dominated fronts and individuals are ranked based on both their Pareto rank and their crowding distance—a measure of solution diversity within a front. Finally, the population is trimmed to its original size, retaining individuals from higher-ranking fronts and preserving diversity. This iterative process continues until convergence criteria are met, yielding an accurate approximation of the Pareto front and enabling efficient exploration of multiobjective optimization spaces.
	
	\section{Results}
	\label{sec:Res}

	In this section, we present the Pareto fronts for each of our models which optimally minimizes the maximum transient second-state population $\max_{t}(p_{\rm{f}})$ in the range $t \in [-T/2, T/2]$ given that we must simultaneously maximize the detuning robustness at the threshold $p_{\rm{e}}(T/2) \ge 0.99$. These fronts are evaluated for pulse durations of $T = 50$ ns and $T = 200$ ns. Since this threshold may be reached at multiple points across the detuning (or amplitude) axis, a consistent definition of robustness is essential. We therefore define detuning robustness (or amplitude robustness) as the width of the $p_{\rm{e}}(T/2)$ curve between the first two points where $p_{\rm{e}}(T/2) = 0.99$, provided that $p_{\rm{e}}(T/2) \geq 0.999$ is reached at least once within this interval. We include amplitude robustness in parenthesis, because even though it was not one of the objectives in our MOO problem, we compute it using the optimal pulse parameters found by the genetic algorithm.

\begin{figure}[h]
		\centering
		\includegraphics[width=1\linewidth]{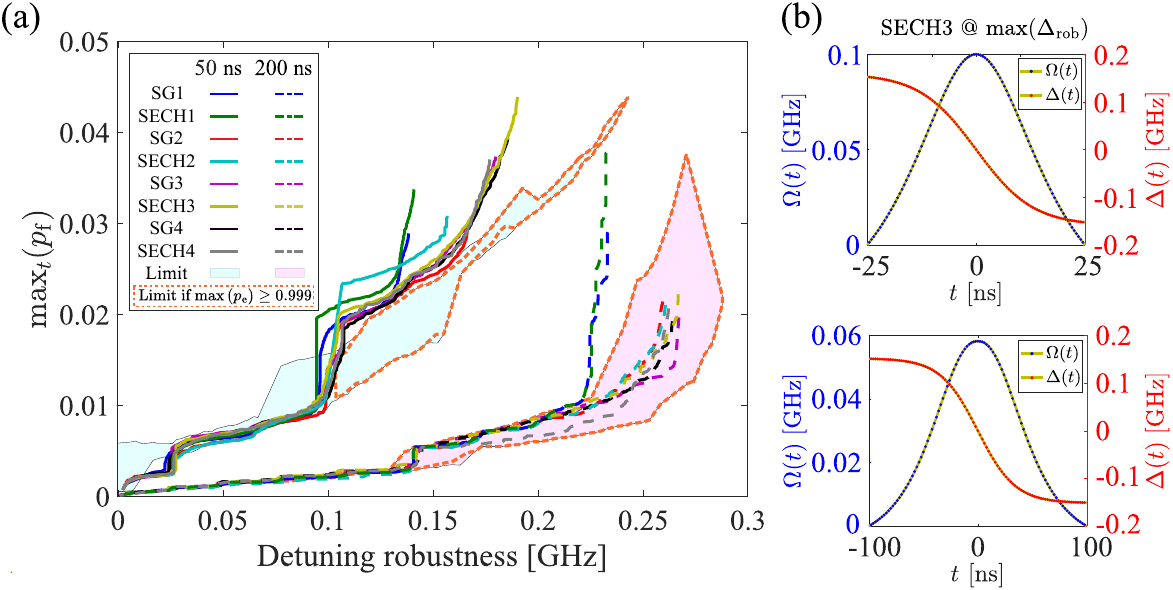}
		\caption{(a) The maximum transient second excited state population $\max_{t}(p_{\rm{f}})$ vs detuning robustness for various models with pulse durations of 50 ns (solid curves) and 200 ns (dashed curves). The shaded regions (cyan and magenta for 50 ns and 200 ns pulses, respectively) represent the range of data constituting the fronts found from using the Landau-Zener (LZ) formula to approximate the theoretical limit of the detuning robustness for each model. These LZ-based bounds are computed at the threshold condition $p_{\rm{e}}^{\rm{thresh}} = 0.99$, but without requiring that $p_{\rm{e}}(T/2) \geq 0.999$ anywhere within the corresponding detuning interval (i.e., between the roots $\delta_{-}$ and $\delta_{+}$, where $p_{\rm{e}}(\delta_{\pm}) = 0.99$). The orange dashed boundary includes this additional constraint. As an example, panel (b) shows the time profiles of the envelope function and detuning function for the SECH3 model at maximum detuning robustness. The upper (lower) tile corresponds to $T = 50$ ($200$) ns.}%
		\label{fig-pareto}
	\end{figure}

	Moreover, the Pareto fronts for each of our models at the two pulse durations $T$ are presented in Fig.~\ref{fig-pareto}(a). Here, we observe that the models form two distinct clusters: one for $T = 50$ ns pulses (solid curves) and one for $T = 200$ ns pulses (dashed curves), with the latter affording us more detuning robustness. For detuning robustness values up to approximately $0.1$ GHz for 50 ns pulses and $0.15$ GHz for 200 ns pulses, all models yield similar $\max_{t}(p_{\rm{f}})$, with the clusters corresponding to respective pulse durations. This suggests that, within these moderate robustness regimes, performance is largely insensitive to the specific choice of pulse model.
	
	We further note that the models with linear detuning functions perform less optimally compared to the other detuning types at higher detuning robustness levels, as evidenced by the divergence towards higher $\max_{t}(p_{\rm{f}})$ values at approximately 0.13 GHz for $50$ ns pulses and at approximately 0.22 GHz for $200$ ns pulses. Otherwise, it is generally less obvious how the other models fare relative to each other. Thus, to better contextualize these results, we assign approximate theoretical bounds to these fronts to ascertain how close we are to the limits imposed by adiabaticity. 
	\\
	To this end, we use the Landau-Zener (LZ) formula in calculating the detuning robustness at the threshold $p_{\rm{e}}(T/2) \geq 0.99$. The LZ formula provides the diabatic transition probability:
	
	\begin{equation}
		p_{\rm LZ} = e^{-2\pi\Gamma},
		\label{eq:LZ}
	\end{equation}
	\newline
	where $\Gamma = \Omega(t_{0})^{2} / 4\hbar |v|$ with $t_{0}$ defined as the time at which the detuning vanishes, $\Delta(t_{0}) = 0$. In the limit $\Gamma \gg 1$, the transition is adiabatic, whereas for $\Gamma \ll 1$, it is nonadiabatic. Here, $v$ is the Landau-Zener velocity, representing the slope of $\Delta(t)$ at the avoided crossing:
	
	\begin{equation}
		v = \left. \frac{d \Delta (t)}{dt} \right|_{t=t_{0}}.
		\label{eq: alpha}
	\end{equation}
	\newline
	For example, in the case of a linear detuning $\Delta(t) = v t + \delta$, we have $t_{0} = -\delta / v$.
	
	 In terms of $v$ and considering the threshold value of the first excited state population we use for calculating the detuning robustness, i.e., $p_{\rm{e}}^{\rm{thresh}} = 0.99$, we impose the condition 
	 
	\begin{equation}
		1 - p_{\rm LZ}\geq p_{\rm{e}}^{\rm{thresh}}\;.
		\label{eq:LZ_thresh}
	\end{equation}
	\newline
	Here, we subtract from unity as Eq.~(\ref{eq:LZ}) describes the probability of remaining in $\vert \rm{g} \rangle$ at the level crossing and we further assume that $\vert \rm{f} \rangle$ is far detuned from the other states (e.g., for $T = 200$ ns pulses, $\vert \rm{f}\rangle$ is only populated if $\delta \approx E_{C}/2$ \cite{Kuzmanovic_2024}). Thus, after substituting $t_{0}$ in Eq.~(\ref{eq:LZ_thresh}) such that it explicitly depends on $\delta$, we solve for $\delta$ to find the two roots of the threshold condition $p_{\rm{e}}^{\rm{thresh}} = 0.99$, denoted $\delta_{-}$ and $\delta_{+}$. Therefore, $\Delta_{\rm{rob}}^{\rm{LZ}} = \delta_{+}-\delta_{-}$. 
	
	These bounds are less stringent than the requirement on the transfer fidelity imposed on the Pareto-optimal solutions, as they do not include the constraint that $p_{\rm{e}}(T/2) \geq 0.999$ anywhere between $\delta_{-}$ and $\delta_{+}$. The corresponding limit that does incorporate this additional constraint is shown as the orange dashed boundary, discernible only for the 50 ns pulses, since for the $200$ ns pulses, the limits with and without this constraint nearly coincide.
		
	Assuming an adiabatic trajectory, $\max_{t}(p_{\rm{f}})$ can be estimated in the following way: given $\Omega(t)$ and $\Delta(t)$ we compute the instantaneous eigenstates of the Hamiltonian (Eq.~(\ref{eq:Ham})) and identify the one which initially corresponds to the ground state $\ket{\psi_{\rm g}(t)}$,  $\ket{\psi_{\rm g}(t=-T/2)}\approx \ket{\rm{g}}$.
	Provided that the evolution is adiabatic, the $\ket{\rm{f}}$ state population as a function of time is given by $p_{\rm f}(t) =|\bra{\rm{f}}\ket{\psi_{\rm g}(t)}|^2$.
	By computing the  maximum with respect to $t$ and combining it with the previously obtained detuning robustness estimate $\Delta_{\rm{rob}}^{\rm{LZ}}$, we obtain the bounds in Fig.~\ref{fig-pareto}(a): the cyan-colored region for the 50 ns pulses and the magenta-colored region for the 200 ns pulses.
	
	Notably, the $50$ ns pulses do not fall within the bounds obtained based on an adiabatic Landau-Zener trajectory, save for the ones with low detuning robustness.  In contrast, the opposite is true for the $200$ ns pulses, where they operate within the adiabatic regime with the exception of the linearly-detuned models at extreme detuning robustness. 
		
	Nonetheless, Fig.~\ref{fig-pareto}(a) reveals that all pulse models are capable of achieving moderate detuning robustness—for instance, $\Delta_{\rm{rob}}=0.1$ GHz—even when the maximum transient population of the second excited state remains low (i.e., $\max_{t}(p_{\rm{f}})$ $\leq 0.01$). Models featuring nonlinear detuning profiles (types 2–4) become essential only when targeting higher levels of detuning robustness. Shorter pulses ($T = 50$ ns) can attain modest robustness but typically incur higher maximum transient second-excited state populations. By contrast, comparable—or in some cases greater—detuning robustness can be achieved with reduced $\max_{t}(p_{\rm{f}})$, albeit at the cost of longer pulse durations ($T = 200$ ns).

	While the Pareto fronts serve as our primary figure of merit, they do not explicitly reveal the values of the Pareto-optimal parameters. In fact, each envelope function has a total of three parameters, i.e., $[ \Omega_{0}, n, \beta\;(\epsilon)]$ for SG (SECH) pulses and depending on the detuning function, there are anywhere from one (for type-1) to five (type-2) detuning parameters. As an example, Fig.~\ref{fig-pareto}(b) shows the envelope and detuning functions at $T = 50$ ns (upper tile) and at $T = 200$ ns (lower tile) for the SECH3 model when its detuning robustness is greatest in our Pareto solution. The parameters which generate these functions for $T = 50$ ($200$) ns are $ \Omega_{0} = 0.1$ ($0.0582$) GHz, $n = 0.995$ ($1.14$), $\epsilon = 2$ ($3.15$), $k_{1} = -0.161$ $(-0.152)$ GHz, and $\gamma_{1} =1.92 $ $(2.92)$. 
	 In the next section, we ascertain the overall importance of each model's parameters with respect to $\max_{t}(p_{\rm{f}})$, detuning robustness $\Delta_{\rm{rob}}$, and amplitude robustness $\Omega_{\rm{rob}}$.

	\section{Parameter Analysis}
	\label{sec:Anal}
	
	\begin{figure}[h]
		\centering
		\includegraphics[width=0.75\linewidth]{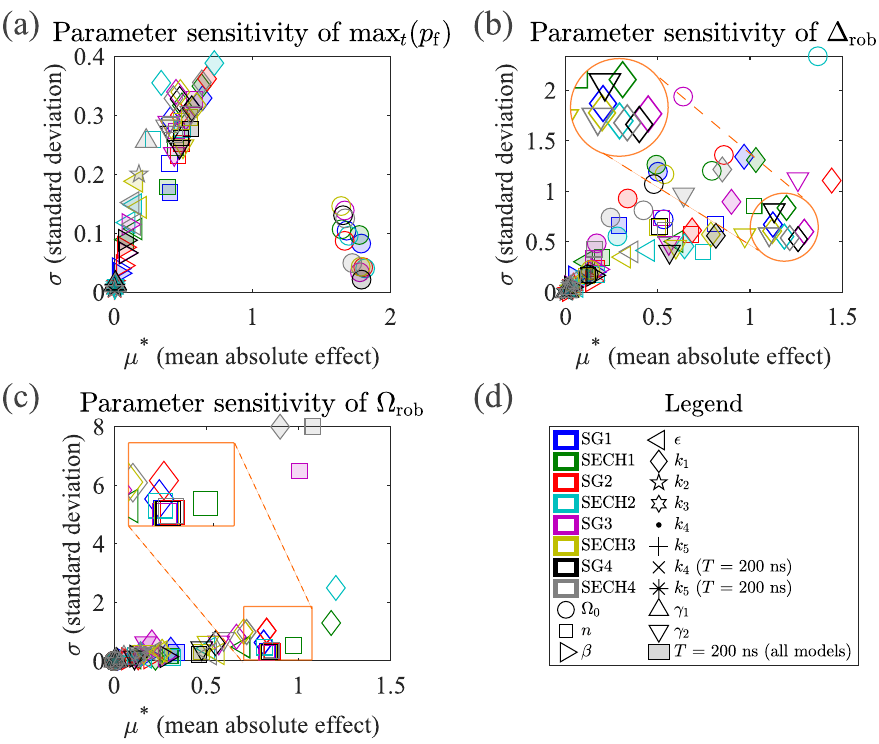}
		\caption{The parameter sensitivity of (a) $\max_{t}(p_{\rm{f}})$, (b) detuning robustness $\Delta_{\rm{rob}}$, and (c) amplitude robustness $\Omega_{\rm{rob}}$, where the vertical axis is the standard deviation $\sigma$ and the horizontal axis is the mean absolute effect $\mu^{*}$ found via the Morris method with a $ 5 \times 10^{-3}\% $ perturbative decrease on each model parameter. The clustering of $k_{1}$ and $n$ is shown with higher resolution in (b) and (c), respectively, via the orange-colored insets. The legend is shown in (d).} %
		\label{fig-params}
	\end{figure}
	
	To assess the impact of the model parameters on each objective outcome, we employ a modified version of the Morris sensitivity test \cite{Morris1991}. This method involves calculating two key metrics: the mean absolute effect $\mu^{*}$ and the standard deviation $\sigma$.
	
	The elementary effect for a given parameter is calculated by perturbing the parameter by a fixed percentage (in our case, a $5 \times 10^{-3}\%$ decrease) while keeping all other parameters fixed. The elementary effect is then defined as the difference between the perturbed and unperturbed model output, divided by the perturbation step size. Unlike the conventional Morris method, we use the relative elementary effect, which is obtained by dividing each elementary effect by the unperturbed output. The mean absolute effect is the average of the absolute values of these relative elementary effects, providing a measure of the overall importance of each parameter in influencing the model output. Explicitly, the mean absolute effect is 
	\begin{equation}
		\mu^{*} = \frac{1}{N}\sum \limits_{i = 1}^{N} \vert E_{i} \vert ,
		\label{eq: mean_abs_effect}
	\end{equation}
	where $E_{i}$ is the relative elementary effect for the $i$-th sample of a parameter, and $N$ is the total number of samples (175 for each of our parameters).
	
	The standard deviation $\sigma$ measures the variability in these relative elementary effects across the parameter space. This quantifies the degree to which interactions between parameters influence the model output. Specifically, we calculate the sample standard deviation of the relative elementary effects for each parameter, i.e., 
	\begin{equation}
		\sigma = \sqrt{\frac{1}{N-1} \sum \limits_{i = 1}^{N} (E_{i} - \mu)^2} ,
		\label{eq: interaction}
	\end{equation}
	where $\mu$ is the mean effect and the $N-1$ in the denominator accounts for the potential variability beyond the analyzed set. Unlike $\mu^{*}$, which represents the mean of the absolute values of the relative elementary effects as defined in Eq.~(\ref{eq: mean_abs_effect}), $\mu$ considers the raw values and retains their signs. The mean effect is a useful metric in its own right, as it indicates whether a parameter's influence on the output is typically positive or negative, revealing the directionality of its impact. Moreover, the ratios of $\mu/\mu^{*}$, which assume values between $-1$ and $1$, inform us of the type of correlation of the parameters relative to the outputs. In Appendix~\ref{sec:Appendix_param_corr}, we have calculated these ratios for $\max_{t}(p_{\rm{f}})$, detuning robustness, and amplitude robustness for each parameter. For instance, we see that for $\max_{t}(p_{\rm{f}})$, the amplitude $\Omega_{0}$ and envelope order $n$ each have a completely positive correlation with $\max_{t}(p_{\rm{f}})$ for all models, i.e., $\mu = \mu^{*}$, so the increase in each of these parameters, keeping all others fixed, yields a higher $\max_{t}(p_{\rm{f}})$. The same can be said for the detuning parameter $k_1$, except for models of type-4 detuning, i.e., SG4 and SECH4 models. For detuning and amplitude robustness, the ratios $\mu/\mu^{*}$ largely lie between $-1$ and $1$ for the models, which suggests a more complex correlation. On the other hand, the envelope parameters $\beta$ and $\epsilon$ are completely negatively correlated, i.e., $\mu = -\mu^{*}$, for all models when the output is $\max_{t}(p_{\rm{f}})$ and nearly completely negatively correlated when the output is detuning robustness. However, for amplitude robustness, these parameters are generally strongly positively correlated for the models. 
	
	The standard deviation quantifies interactions among parameters, as the effect of changing a parameter will vary based on the levels of other parameters if strong interactions are present. This variability causes a higher spread in $E_{i}$, resulting in a larger $\sigma$. Moreover, if increasing one parameter has a different impact depending on whether another parameter is high or low, this interaction will increase the variability of $E_{i}$ for the former parameter, and consequently, $\sigma$ for that parameter will be large. 
	
	While $\mu^*$ measures the overall importance of a parameter (how much it influences the output on average), $\sigma$ highlights whether that influence is consistent or varies significantly.
	A parameter with high $\mu^{*}$ but low $\sigma$ suggests a dominant, consistent effect, while high $\mu^*$ and high $\sigma$ suggest strong influence with significant variability, likely due to interactions or nonlinearity. 
	
	As shown in Fig.~\ref{fig-params}(a), there is a clear cluster of $\Omega_0$ parameters with high mean absolute effect ($\mu^{*}$) and low standard deviation ($\sigma $) across all models and pulse durations. This indicates that $\Omega_{0}$ has the strongest and most consistent influence on minimizing the maximum transient second excited state population $\max_t(p_{\rm f})$.

	In contrast, Fig.~\ref{fig-params}(b) and (c) display generally higher variability among parameters, indicating a more complex interplay. However, Fig.~\ref{fig-params}(b) reveals a prominent high-$\mu^{*}$ cluster for $k_1$, suggesting that $k_1$ plays a dominant role in determining detuning robustness $\Delta_{\rm rob}$—an aspect analyzed further in Appendix~\ref{sec:Appendix_detrob}. Amplitude robustness, in particular how it relates to detuning robustness, is discussed in Appendix~\ref{sec:Appendix_amprob}. We note in Fig.~\ref{fig-params}(c) that there is a noticeable cluster of envelope order $n$, suggesting it is an impactful parameter in determining amplitude robustness.

	\section{Discussion and conclusions}
	\label{sec:Con}

	Inspired by exactly solvable two-state models, we have investigated the Pareto-optimality of population transfer from the ground state to the first excited state in a ladder-type qutrit. Our optimization considered two competing objectives: minimizing the maximum transient population of the second excited state and maximizing detuning robustness. We define the latter as the width of the first excited state population $p_{\rm{e}}(T/2)$ curve between the first two points where $p_{\rm{e}}(T/2) = 0.99$, provided that $p_{\rm{e}}(T/2) \geq 0.999$ is reached at least once within this interval. Due to the inherent trade-off between these objectives, the Pareto front represents the set of optimal solutions to this multiobjective problem.
	
	We explored eight pulse models, combining two envelope functions (super-Gaussian or hyperbolic secant) with four detuning profiles: linear (type-1), quintic (type-2), hyperbolic tangent (type-3), and a superposition of hyperbolic secant and hyperbolic tangent (type-4). All pulses satisfy $\Omega(0) = \Omega_{0}$, $\Omega(\pm T/2) = 0$, and $\Delta(0) - \delta = 0$. We optimized parameters for two pulse durations, $T = 50$ ns and $T = 200$ ns. 
	
 	The Pareto fronts, shown in Fig.~\ref{fig-pareto}(a), reveal two clusters based on pulse duration, with longer pulses yielding superior trade-offs. Comparing with Landau-Zener-based bounds derived from local linearization of detuning functions, we find that the $T = 50$ ns pulses lie outside the adiabatic regime, especially at higher values of detuning robustness. Among the $T = 200$ ns pulses, only the models with nonlinear detunings (types 2--4) lie inside the Landau-Zener bounds when delivering high robustness where the models with type-3 and type-4 detunings perform best. As such, our results suggest that detuning shape has a greater impact than envelope choice.
	
	All pulse models perform similarly at moderate detuning robustness—up to about 0.1 GHz (50 ns) and 0.15 GHz (200 ns)—with low transient excitation (i.e., $\max_{t}(p_{\rm{f}}) \leq 0.01$) in all cases, as shown by the Pareto fronts. However, achieving higher detuning robustness generally requires nonlinear detuning profiles (types 2–4). Longer pulses also tend to improve performance by further reducing transient excitation.
	
	Finally, we applied the Morris method to assess parameter sensitivity. The most influential parameter for $\max_{t}(p_{\rm{f}})$ is the amplitude $\Omega_{0}$, consistently showing high mean absolute effect and low standard deviation. For detuning and amplitude robustness, influence is generally more distributed, with the exception of the detuning prefactor $k_{1}$, which clearly dominates for detuning robustness.

	In conclusion, our study provides a systematic approach for designing optimal pulses that enable robust population transfer in ladder-type qutrits. The insights gained here are broadly applicable to any system where the pulse models discussed can be implemented.

\section*{Declarations}

\bmhead{Funding} 	

This project has received funding from the EU Flagship on Quantum Technology through HORIZON-CL4-2022-QUANTUM-01-SGA Project No. 101113946 OpenSuperQPlus100.\;\;We also acknowledge support from Research Council of Finland Proof-of-Concept program (Project 359397) and from the Centre of Excellence program (Project 352925).

\bmhead{Competing interests} 
The authors declare no competing interests.

\bmhead{Ethics approval and consent to participate} 
Not applicable.

\bmhead{Consent for publication} 
Not applicable.

\bmhead{Data Availability} 
Data is available upon reasonable request.

\bmhead{Materials availability} 
Not applicable.

\bmhead{Code availability}
Codes are available upon reasonable request.

\bmhead{Author contribution}
J.J.M. performed the numerical experiments and data processing. J.J.M. and M.K. developed the methodology and performed the analysis. M.K. proposed the initial research idea and together with G.S.P. supervised the project. All authors contributed to writing of the manuscript.\\

%
%
%
%
%
%

\begin{appendices}
	\counterwithin{figure}{section}
\counterwithin{table}{section}
\setcounter{table}{0}
\section{Parameter correlations with the outputs}
\label{sec:Appendix_param_corr}
Following the Morris method as described in Sect.~\ref{sec:Anal}, we have tabulated $\mu/\mu^{*}$ for outputs $\max_{t}(p_{\rm{f}})$, $\Delta_{\rm{rob}}$, and $\Omega_{\rm{rob}}$ in Tables~\ref{tab:mu_ratios_max_pf}, \ref{tab:det_rob}, \ref{tab:amp_rob}, respectively. These ratios necessarily range from $-1$ to $1$, where a ratio of $-1$ indicates a parameter is completely negatively correlated with the output, and a ratio of $1$ indicates a parameter is completely positively correlated with the output. 
\subsection{Parameter correlations with respect to $\max_{t}(p_{\rm{f}})$}
We see in Table~\ref{tab:mu_ratios_max_pf} that for $\max_{t}(p_{\rm{f}})$, the ratios of $\mu/\mu^{*}$ corresponding to the envelope parameters $\Omega_{0}$ and $n$ are unity for each model and pulse duration, suggesting that they are completely positively correlated with output. However, for the other envelope parameter $\beta \backslash \epsilon$ the opposite is true; i.e., the ratios are all negatively correlated with $\max_{t}(p_{\rm{f}})$. Among the detuning parameters, $k_{1}$ and $\gamma_{1}$ are both completely positively correlated; whereas the results for the other parameters are more mixed, with the notable exception of $k_{5}$, where the ratios suggest strong negative correlations with the objective. It is also worth emphasizing that the parameter $k_{2}$ for the type-2 detuning models and the type-4 detuning functions are fundamentally different: in the former, $k_{2}$ controls the convexity or concavity of the detuning function and in the latter, $k_{2}$ determines the size of the dip which the hyperbolic secant term introduces in the detuning function. Thus, this discrepancy in the functionality of this parameter among the type-2 and type-4 detuning functions is reflected in the more extreme values of the ratios for the models with type-2 detuning compared to those of type-4, with the exception of the SECH4 ($T = 200$ ns) model, where the ratios are interestingly $-1$. In order to meet our Pareto objective which involves the minimization of $\max_{t}(p_{\rm{f}})$ over $t \in [-T/2, T/2]$, Table~\ref{tab:mu_ratios_max_pf} informs us that we should avoid high values of $\Omega_{0}$, $n$, $k_{1}$, and $\gamma_{1}$, while seeking more extreme values, i.e., further from zero, of $\beta \backslash \epsilon$ and $k_{5}$.

\begin{table*}[h]
	\renewcommand\arraystretch{1}
	\centering
	\caption{Ratios of $\mu / \mu^*$ for $\max_{t}(p_{\rm{f}})$}
	\tiny 
	\begin{tabular}{|l|c|c|c|c|c|c|c|c|c|c|}	
		\hline
		\textbf{Model ($T$)} & \textbf{$\Omega_0$} & \textbf{$n$} &  \textbf{$\beta \backslash \epsilon$} & \textbf{$k_1$} & \textbf{$k_2$} & \textbf{$k_3$} & \textbf{$k_4$} & \textbf{$k_5$} & \textbf{$\gamma_1$} & \textbf{$\gamma_2$}\\
		\hline
		SG1 (50 ns) & \colorvalue{1}  & \colorvalue{1}  & \colorvalue{-1}  & \colorvalue{1} & - & -  & -  & -  & - & - \\
		SG1 (200 ns) & \colorvalue{1}  & \colorvalue{1}  & \colorvalue{-1}  & \colorvalue{1} & - & -  & -  & -  & - & - \\
		SECH1 (50 ns)  & \colorvalue{1}  & \colorvalue{1}  & \colorvalue{-1}  & \colorvalue{1} & - & -  & -  & -  & - & -  \\
		SECH1 (200 ns)  & \colorvalue{1}  & \colorvalue{1}  & \colorvalue{-1}  & \colorvalue{1} & - & - & -  & -  & - & - \\
		SG2 (50 ns)  & \colorvalue{1}  & \colorvalue{1}  & \colorvalue{-1}  & \colorvalue{1} & \colorvalue{0.91} & \colorvalue{-0.81}  & \colorvalue{-0.77}  & \colorvalue{-0.93}  & - & - \\
		SG2 (200 ns)  & \colorvalue{1}  & \colorvalue{1}  & \colorvalue{-1}  & \colorvalue{1} & \colorvalue{-0.99} & \colorvalue{-0.91} & \colorvalue{0.2}  & \colorvalue{-0.99}  & - & - \\
		SECH2 (50 ns)  & \colorvalue{1}  & \colorvalue{1}  & \colorvalue{-1}  & \colorvalue{1} & \colorvalue{-0.97} & \colorvalue{0.95}  & \colorvalue{0.38}  & \colorvalue{-0.91}  & - & - \\
		SECH2 (200 ns)  & \colorvalue{1}  & \colorvalue{1}  & \colorvalue{-1}  & \colorvalue{1} & \colorvalue{-0.95} & \colorvalue{-0.36}  & \colorvalue{-0.65}  & \colorvalue{-0.94}  & - & - \\
		SG3 (50 ns)  & \colorvalue{1}  & \colorvalue{1}  & \colorvalue{-0.99}  & \colorvalue{1} & - & -  & -  & -  & \colorvalue{1} & - \\
		SG3 (200 ns)  & \colorvalue{1}  & \colorvalue{1}  & \colorvalue{-1}  & \colorvalue{1} & - & -  & -  & -  & \colorvalue{1} & -  \\
		SECH3 (50 ns)  & \colorvalue{1}  & \colorvalue{1}  & \colorvalue{-1}  & \colorvalue{1} & - & -  & -  & -  & \colorvalue{1} & - \\
		SECH3 (200 ns)  & \colorvalue{1}  & \colorvalue{1}  & \colorvalue{-1}  & \colorvalue{1} & - & -  & -  & -  & \colorvalue{1} & - \\
		SG4 (50 ns)  & \colorvalue{1}  & \colorvalue{1}  & \colorvalue{-0.99}  & \colorvalue{1} & \colorvalue{0.01} & -  & -  & -  & \colorvalue{1} & \colorvalue{0.057}  \\
		SG4 (200 ns)  & \colorvalue{1}  & \colorvalue{1}  & \colorvalue{-1}  & \colorvalue{1} & \colorvalue{-0.46} & -  & -  & -  & \colorvalue{1} & \colorvalue{-0.46}  \\
		SECH4 (50 ns)  & \colorvalue{1}  & \colorvalue{1}  & \colorvalue{-1}  & \colorvalue{1} & \colorvalue{0.047} & -  & -  & -  & \colorvalue{1} & \colorvalue{0.057} \\
		SECH4 (200 ns)  & \colorvalue{1}  & \colorvalue{1}  & \colorvalue{-1}  & \colorvalue{1} & \colorvalue{-1} & -  & -  & -  & \colorvalue{1} & \colorvalue{-1} \\
		\hline
	\end{tabular}
	
	\setcounter{table}{0}
	\caption{The ratios of $\mu/\mu^{*}$ for $\max_{t}(p_{\rm{f}})$ for each model and pulse duration when each parameter is independently perturbed by a $5 \times 10^{-3}\%$ decrease following the Morris method.  Positive ratios (red) indicate an overall positive correlation and negative ratios (blue) indicate an overall negative correlation.}
	\label{tab:mu_ratios_max_pf}
\end{table*}

\subsection{Parameter correlations with respect to $\Delta_{\rm{rob}}$}
The $\mu/\mu^{*}$ ratios for detuning robustness $\Delta_{\rm{rob}}$ are tabulated in Table~\ref{tab:det_rob} for each of the models. For the envelope parameters, we again see the same trend as in Table~\ref{tab:mu_ratios_max_pf}, though with generally less extreme values. We note, for instance, that the ratios for $\Omega_{0}$ are all such that $ 0.5 \leq \mu/\mu^{*} \leq 0.97$. For the parameter $n$, the ratios generally suggest a stronger positive correlation than for $\Omega_{0}$, and the ratios for $\beta \backslash \epsilon$ are all negative, but with less extreme values than in Table~\ref{tab:mu_ratios_max_pf}. Further, we note that the ratios corresponding to $n$ and $\beta \backslash \epsilon$ appear to be less extreme for the longer pulse duration, suggesting that a longer pulse duration for these parameters yields inconsistent directional effects. The ratios corresponding to each of the detuning parameters are also less extreme in Table~\ref{tab:det_rob}, with the $\mu/\mu^{*}$ ratios of $k_{1}$ and $\gamma_{1}$ suggesting strong positive correlation with $\Delta_{\rm{rob}}$. For the other parameters, the ratios generally suggest that the models have inconsistent effects relative to each other. 

\begin{table*}[h]
	\renewcommand\arraystretch{1}
	\centering
	\caption{Ratios of $\mu / \mu^*$ for $\Delta_{\rm{rob}}$}
	\tiny 
	\begin{tabular}{|l|c|c|c|c|c|c|c|c|c|c|}
		\hline
		\textbf{Model ($T$)} & \textbf{$\Omega_0$} & \textbf{$n$} &  \textbf{$\beta \backslash \epsilon$} & \textbf{$k_1$} & \textbf{$k_2$} & \textbf{$k_3$} & \textbf{$k_4$} & \textbf{$k_5$} & \textbf{$\gamma_1$} & \textbf{$\gamma_2$}\\
		\hline
		SG1 (50 ns) & \colorvalue{0.76}  & \colorvalue{1}  & \colorvalue{-1}  & \colorvalue{0.98} & - & -  & -  & -  & - & - \\
		SG1 (200 ns) & \colorvalue{0.7}  & \colorvalue{0.85}  & \colorvalue{-0.82}  & \colorvalue{0.53} & - & -  & -  & -  & - & - \\
		SECH1 (50 ns)  & \colorvalue{0.95}  & \colorvalue{1}  & \colorvalue{-1}  & \colorvalue{0.93} & - & -  & -  & -  & - & -  \\
		SECH1 (200 ns)  & \colorvalue{0.69}  & \colorvalue{0.61}  & \colorvalue{-0.63}  & \colorvalue{0.5} & - & - & -  & -  & - & - \\
		SG2 (50 ns)  & \colorvalue{0.88}  & \colorvalue{1}  & \colorvalue{-1}  & \colorvalue{0.99} & \colorvalue{0.57} & \colorvalue{-0.56}  & \colorvalue{-0.82}  & \colorvalue{0.13}  & - & - \\
		SG2 (200 ns)  & \colorvalue{0.93}  & \colorvalue{0.94}  & \colorvalue{-0.83}  & \colorvalue{0.83} & \colorvalue{-0.4} & \colorvalue{-0.41} & \colorvalue{0.045}  & \colorvalue{0.084}  & - & - \\
		SECH2 (50 ns)  & \colorvalue{0.97}  & \colorvalue{0.99}  & \colorvalue{-1}  & \colorvalue{0.98} & \colorvalue{-0.99} & \colorvalue{-0.75}  & \colorvalue{0.55}  & \colorvalue{-0.44}  & - & - \\
		SECH2 (200 ns)  & \colorvalue{0.71}  & \colorvalue{0.94}  & \colorvalue{-0.96}  & \colorvalue{0.96} & \colorvalue{-0.7} & \colorvalue{0.55}  & \colorvalue{-0.047}  & \colorvalue{-0.28}  & - & - \\
		SG3 (50 ns)  & \colorvalue{0.6}  & \colorvalue{0.95}  & \colorvalue{-1}  & \colorvalue{0.99} & - & -  & -  & -  & \colorvalue{1} & - \\
		SG3 (200 ns)  & \colorvalue{0.5}  & \colorvalue{0.92}  & \colorvalue{-0.88}  & \colorvalue{0.84} & - & -  & -  & -  & \colorvalue{0.99} & -  \\
		SECH3 (50 ns)  & \colorvalue{0.91}  & \colorvalue{0.97}  & \colorvalue{-0.99}  & \colorvalue{0.97} & - & -  & -  & -  & \colorvalue{1} & - \\
		SECH3 (200 ns)  & \colorvalue{0.69}  & \colorvalue{0.91}  & \colorvalue{-0.96}  & \colorvalue{0.92} & - & -  & -  & -  & \colorvalue{0.97} & - \\
		SG4 (50 ns)  & \colorvalue{0.81}  & \colorvalue{0.96}  & \colorvalue{-1}  & \colorvalue{1} & \colorvalue{-0.81} & -  & -  & -  & \colorvalue{0.99} & \colorvalue{-0.81}  \\
		SG4 (200 ns)  & \colorvalue{0.79}  & \colorvalue{0.97}  & \colorvalue{-0.97}  & \colorvalue{0.92} & \colorvalue{-0.5} & -  & -  & -  & \colorvalue{0.97} & \colorvalue{-0.7}  \\
		SECH4 (50 ns)  & \colorvalue{0.81}  & \colorvalue{0.98}  & \colorvalue{-1}  & \colorvalue{0.99} & \colorvalue{-0.38} & -  & -  & -  & \colorvalue{0.99} & \colorvalue{-0.31} \\
		SECH4 (200 ns)  & \colorvalue{0.55}  & \colorvalue{0.71}  & \colorvalue{-0.75}  & \colorvalue{0.76} & \colorvalue{-0.38} & -  & -  & -  & \colorvalue{0.78} & \colorvalue{0.55} \\
		\hline
	\end{tabular}
	\setcounter{table}{1}
	\caption{The ratios of $\mu/\mu^{*}$ for $\Delta_{\rm{rob}}$ for each model and pulse duration when each parameter is independently perturbed by a $5 \times 10^{-3}\%$ decrease following the Morris method. Positive ratios (red) indicate an overall positive correlation and negative ratios (blue) indicate an overall negative correlation.}
	\label{tab:det_rob}
\end{table*}

\subsection{Parameter correlations with respect to $\Omega_{\rm{rob}}$}
Table~\ref{tab:amp_rob} shows the $\mu/\mu^{*}$ ratios for amplitude robustness $\Omega_{\rm{rob}}$. The ratios suggest generally weak and mixed directional effects for the amplitude parameter $\Omega_{0}$, which is not at all surprising as $\Omega_{0}$ principally has no effect on amplitude robustness. Further, we observe the ratios for the parameters $n$ and $\beta \backslash \epsilon$ indicate that they are generally strongly negatively correlated and positively correlated with $\Omega_{\rm{rob}}$, respectively. This correlation is opposite to what was observed for $\max_{t}(p_{\rm{f}})$ and $\Delta_{\rm{rob}}$. Across most models, $\mu/\mu^{*}$ of $k_{1}$ suggests negative correlation with amplitude robustness to varying degrees. The exceptions are for the SG4 ($200$ ns) and SECH4 ($200$ ns) models, where $k_{1}$ has the effect of scaling the detuning function, rather than determining its slope as in models of type-1 and type-2 detuning functions. Finally, we note that the ratios corresponding to $\gamma_{1}$ are indicative of a negative effect on $\Omega_{\rm{rob}}$. For the other parameters, the ratios generally suggest that the models have inconsistent effects relative to each other. 

\begin{table*}[h]
	\renewcommand\arraystretch{1}
	\centering
	\caption{Ratios of $\mu / \mu^*$ for $\Omega_{\rm{rob}}$}
	\tiny 
	\begin{tabular}{|l|c|c|c|c|c|c|c|c|c|c|}
		\hline
		\textbf{Model ($T$)} & \textbf{$\Omega_0$} & \textbf{$n$} &  \textbf{$\beta \backslash \epsilon$} & \textbf{$k_1$} & \textbf{$k_2$} & \textbf{$k_3$} & \textbf{$k_4$} & \textbf{$k_5$} & \textbf{$\gamma_1$} & \textbf{$\gamma_2$}\\
		\hline
		SG1 (50 ns) & \colorvalue{0.2}  & \colorvalue{-1}  & \colorvalue{0.99}  & \colorvalue{-0.79} & - & -  & -  & -  & - & - \\
		SG1 (200 ns) & \colorvalue{0.42}  & \colorvalue{-1}  & \colorvalue{0.98}  & \colorvalue{-0.81} & - & -  & -  & -  & - & - \\
		SECH1 (50 ns)  & \colorvalue{-0.19}  & \colorvalue{-1}  & \colorvalue{0.97}  & \colorvalue{-0.91} & - & -  & -  & -  & - & -  \\
		SECH1 (200 ns)  & \colorvalue{-0.087}  & \colorvalue{-1}  & \colorvalue{1}  & \colorvalue{-0.99} & - & - & -  & -  & - & - \\
		SG2 (50 ns)  & \colorvalue{-0.044}  & \colorvalue{-1}  & \colorvalue{0.99}  & \colorvalue{-0.47} & \colorvalue{-0.23} & \colorvalue{0.18}  & \colorvalue{-0.57}  & \colorvalue{-0.44}  & - & - \\
		SG2 (200 ns)  & \colorvalue{-0.1}  & \colorvalue{-1}  & \colorvalue{0.99}  & \colorvalue{-0.89} & \colorvalue{-0.44} & \colorvalue{-0.21} & \colorvalue{0.044}  & \colorvalue{-0.21}  & - & - \\
		SECH2 (50 ns)  & \colorvalue{0.2}  & \colorvalue{-0.98}  & \colorvalue{0.9}  & \colorvalue{-0.67} & \colorvalue{-0.76} & \colorvalue{0.72}  & \colorvalue{-0.074}  & \colorvalue{0.65}  & - & - \\
		SECH2 (200 ns)  & \colorvalue{0.23}  & \colorvalue{-1}  & \colorvalue{1}  & \colorvalue{-0.87} & \colorvalue{-0.58} & \colorvalue{-0.3}  & \colorvalue{-0.83}  & \colorvalue{0.14}  & - & - \\
		SG3 (50 ns)  & \colorvalue{-0.08}  & \colorvalue{-0.99}  & \colorvalue{0.99}  & \colorvalue{-0.67} & - & -  & -  & -  & \colorvalue{-0.79} & - \\
		SG3 (200 ns)  & \colorvalue{0.85}  & \colorvalue{-1}  & \colorvalue{1}  & \colorvalue{-0.74} & - & -  & -  & -  & \colorvalue{-0.98} & -  \\
		SECH3 (50 ns)  & \colorvalue{0.067}  & \colorvalue{-1}  & \colorvalue{1}  & \colorvalue{-0.43} & - & -  & -  & -  & \colorvalue{-0.56} & - \\
		SECH3 (200 ns)  & \colorvalue{-0.23}  & \colorvalue{-1}  & \colorvalue{1}  & \colorvalue{-0.97} & - & -  & -  & -  & \colorvalue{-0.99} & - \\
		SG4 (50 ns)  & \colorvalue{0.14}  & \colorvalue{-0.99}  & \colorvalue{1}  & \colorvalue{-0.56} & \colorvalue{-0.89} & -  & -  & -  & \colorvalue{-0.78} & \colorvalue{-0.87}  \\
		SG4 (200 ns)  & \colorvalue{-0.052}  & \colorvalue{-0.96}  & \colorvalue{0.89}  & \colorvalue{0.9} & \colorvalue{0.33} & -  & -  & -  & \colorvalue{-0.97} & \colorvalue{0.3}  \\
		SECH4 (50 ns)  & \colorvalue{0.28}  & \colorvalue{-0.99}  & \colorvalue{0.99}  & \colorvalue{-0.47} & \colorvalue{-0.82} & -  & -  & -  & \colorvalue{-0.61} & \colorvalue{-0.78} \\
		SECH4 (200 ns)  & \colorvalue{-0.25}  & \colorvalue{0.13}  & \colorvalue{1}  & \colorvalue{0.67} & \colorvalue{-1} & -  & -  & -  & \colorvalue{-0.32} & \colorvalue{-0.98} \\
		\hline
	\end{tabular}
	\setcounter{table}{2}
	\caption{The ratios of $\mu/\mu^{*}$ for $\Omega_{\rm{rob}}$ for each model and pulse duration when each parameter is independently perturbed by a $5 \times 10^{-3}\%$ decrease following the Morris method. Positive ratios (red) indicate an overall positive correlation and negative ratios (blue) indicate an overall negative correlation.}
	\label{tab:amp_rob}
\end{table*}

\clearpage

\section{Parameter Effects on the Pareto Objectives}
\label{sec:Appendix_detrob}

This Appendix analyzes how the Pareto-optimal amplitude $\Omega_{0}$ and detuning prefactor $k_{1}$ influence the two Pareto objectives: the maximum transient second excited state population, $\max_{t}(p_{\rm{f}})$, which we aim to minimize, and the detuning robustness, $\Delta_{\rm rob}$, which we seek to maximize.
\subsection{Effect of $\Omega_{0}$ on the Pareto Objectives}
\begin{figure}[H]
	\centering
	\includegraphics[width=0.5\linewidth]{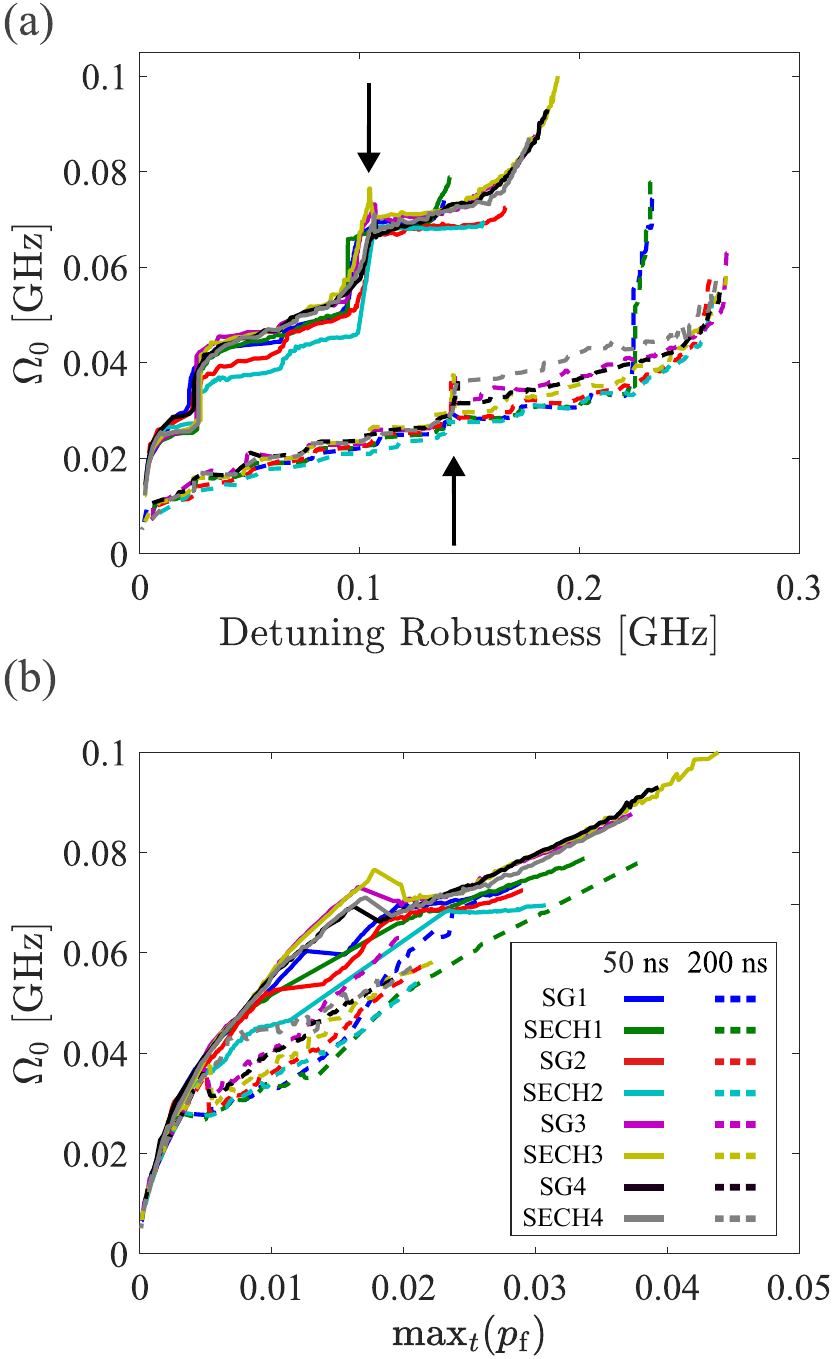}
	\caption{The Pareto-optimal amplitudes $\Omega_{0}$ vs detuning robustness in panel (a) and $\max_{t}(p_{\rm{f}})$ in panel (b). Each model is evaluated at pulse durations $T = 50$ ns (solid lines) and $T = 200$ ns (dashed lines). The vertical arrows in (a) indicate that there are jumps in $\Omega_{0}$ corresponding to a change in sign of $k_{1}$.}%
	\label{fig-amp_vs_max_pf_and_detrob}
\end{figure}

\FloatBarrier
As observed in Sect.~\ref{sec:Res}, longer pulses offer greater robustness to detuning. This explains the distinct clustering seen in Fig.~\ref{fig-amp_vs_max_pf_and_detrob}(a), where the Pareto-optimal amplitude is plotted against detuning robustness for both the 50 ns and 200 ns pulse durations of each model. Notably, the 50 ns pulses exhibit higher values of $\Omega_{0}$ than the 200 ns pulses at the same level of detuning robustness.

In Fig.~\ref{fig-amp_vs_max_pf_and_detrob}(b), the Pareto-optimal amplitudes $\Omega_{0}$ for each model and pulse duration approach zero as $\max_{t}(p_{\rm{f}}) \rightarrow 0$. As $\max_{t}(p_{\rm{f}})$ increases, the amplitudes corresponding to shorter pulses ($T = 50$ ns) tend to be higher than those for longer pulses ($T = 200$ ns). 
Overall, a strong correlation is evident: $\Omega_0$ increases nearly monotonically with $\max_t(p_{\rm{f}})$, forming a universal curve that is independent of the pulse shape, duration, or detuning function.
This trend is corroborated by the Morris analysis in Fig.~\ref{fig-params}(a), where $\Omega_{0}$ shows a high impact—characterized by a large mean absolute effect and low standard deviation---relative to the other parameters.

\subsection{Effect of $k_{1}$ on the Pareto objectives}
\FloatBarrier

\begin{figure}[h]
\centering
\includegraphics[width=0.65\linewidth]{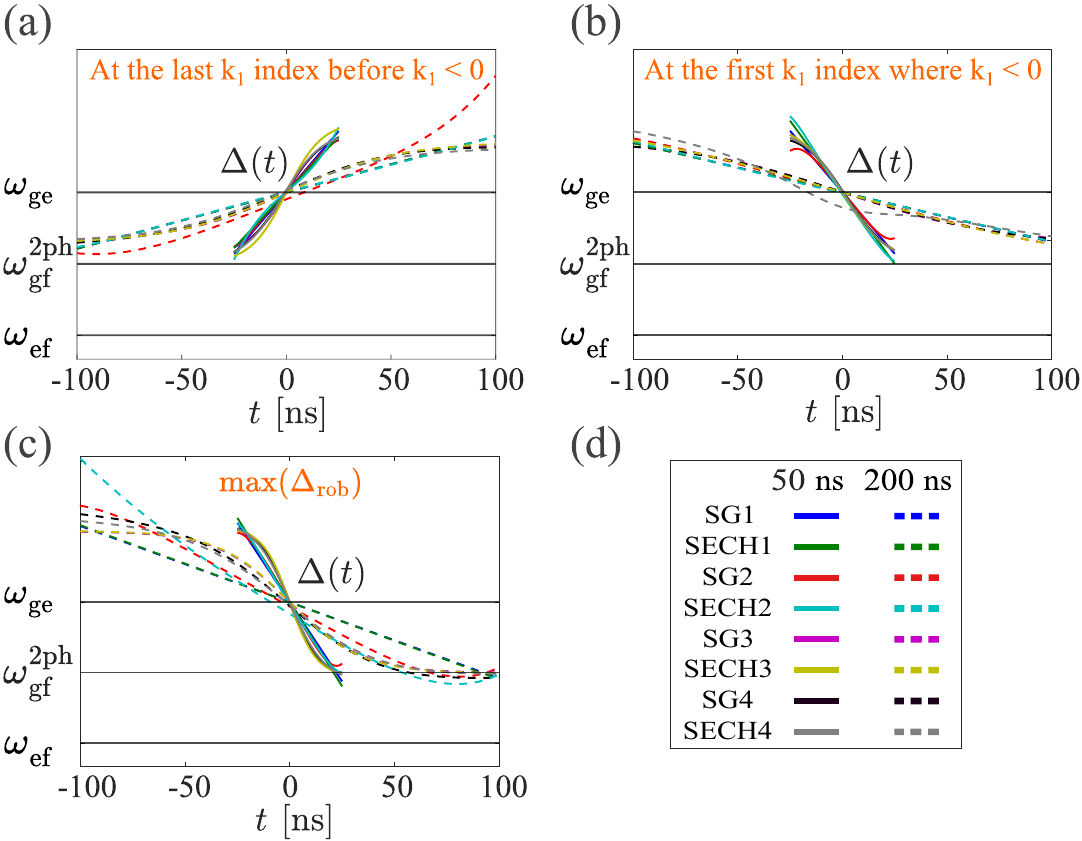}
\caption{The Pareto-optimal detuning functions for each model at pulse durations $T = 50$ ns (solid lines) and $T = 200$ ns (dashed lines). Panel (a) shows the Pareto-optimal $\Delta(t)$ at the last positive $k_{1}$ index, whereas (b) shows the Pareto-optimal $\Delta(t)$ at the first negative $k_{1}$ index. Panel (c) shows the most detuning robust Pareto-optimal detuning function and (d) contains the legend.}%
\label{fig-detuning_functions}
\end{figure}

Here, we investigate how different detuning profiles influence the Pareto objectives, with particular focus on the parameter $k_1$. As demonstrated by the Morris analysis in the main text, $k_1$ has the most significant impact. Moreover, as will be shown shortly, the detuning profiles are approximately linear---especially near the resonance point $\Delta(t) = 0$---with higher-order terms contributing only minor corrections (cf. Fig.~\ref{fig-detuning_functions}).

\begin{figure}[h!]
	\centering
	\includegraphics[width=0.65\linewidth]{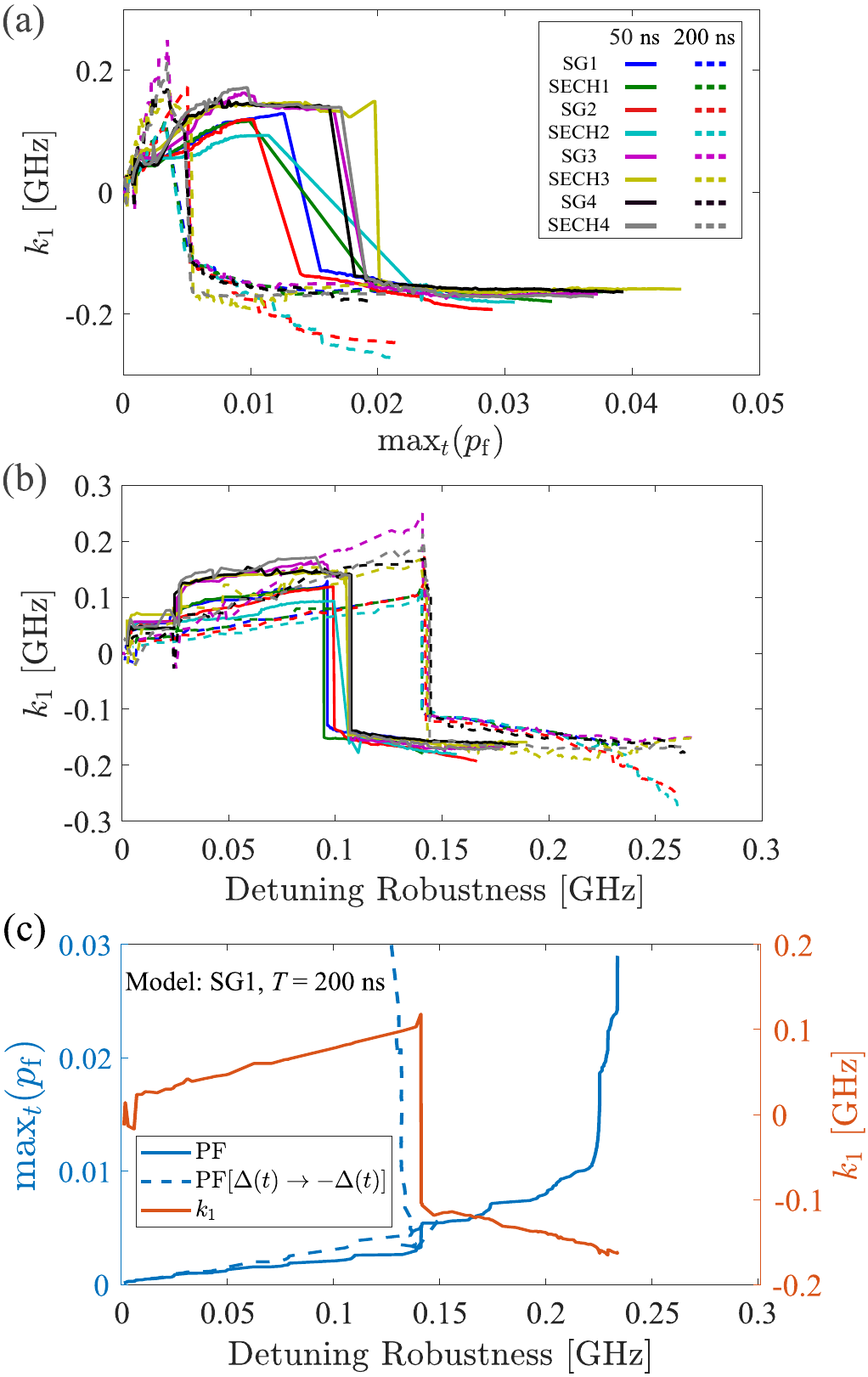}
	\caption{The Pareto-optimal detuning prefactors $k_{1}$ as a function of (a) $\max_{t}(p_{\rm{f}})$ and (b) detuning robustness for each model at pulse durations $T = 50$ ns (solid lines) and $T = 200$ ns (dashed lines). Panel (c) shows the Pareto front (PF) with our usual definition of $\Delta(t)$ (solid blue line) and the Pareto front with $\Delta(t) \rightarrow -\Delta(t)$ (blue dotted line) for the SG1 model with a pulse duration of $T = 200$ ns. The right vertical axis shows the values of the corresponding Pareto-optimal $k_{1}$ (solid red line).} 
\label{fig-k1_and_max_detrob}
\end{figure}

As in the previous section, we begin by plotting the $k_{1}$ parameter as a function of the Pareto objectives, shown in Fig.~\ref{fig-k1_and_max_detrob}. There we observe that, opposite of what was found for $\Omega_{0}$, $k_1$ is highly correlated with the detuning robustness, and much less so with $\max_{t}(p_{\rm{f}})$. Namely, as a function of the detuning robustness, $k_1$ traces cluster into two universal curves, one for each of the pulse durations. Moreover, we find that $|k_{1}|$ is an approximately linear function of $\Delta_{\rm rob}$, with a sign switch occurring mid-way. The same discontinuity can be observed in the Pareto fronts: cf. Fig.~\ref{fig-pareto}(a) of the main text or Fig.~\ref{fig-amp_vs_max_pf_and_detrob} where the discontinuities are marked with arrows.

To understand the origin of the discontinuity it is instructive to plot the detuning profiles on top of the qutrit spectrum. This is shown in Fig.~\ref{fig-detuning_functions} for several different levels of detuning robustness. We observe that for small values of detuning robustness $k_1>0$, as we approach the $\mathrm{g-e}$ transition frequency from below (panel a). However, for larger values of $k_1$, $\Delta(t)$ will become nearly resonant with the two-photon $\mathrm{g-f}$ transition close to the beginning of the pulse which will inherently populate the second excited state. To avoid this, the sign of the detuning changes (see panels b and c), and the 2-photon resonance is driven only after the $\mathrm{g-e}$ transfer is complete, thus having little effect. To complete the analysis, we evaluate the Pareto front with a sign change of the detuning function $\Delta(t) \rightarrow -\Delta(t)$, from which we find that for small values of $k_1$/$\Delta_{\rm{rob}}$ this leads to an increase of $\max_{t}(p_{\rm{f}})$; thus requiring a sign change of $k_1$ to achieve the optimal performance. This is illustrated in Fig.~\ref{fig-k1_and_max_detrob}(c) for the SG1 model with a pulse duration of $T = 200$ ns, which is represented as a blue dotted line.  

\begin{figure}[h]
\centering
\includegraphics[width=0.65\linewidth]{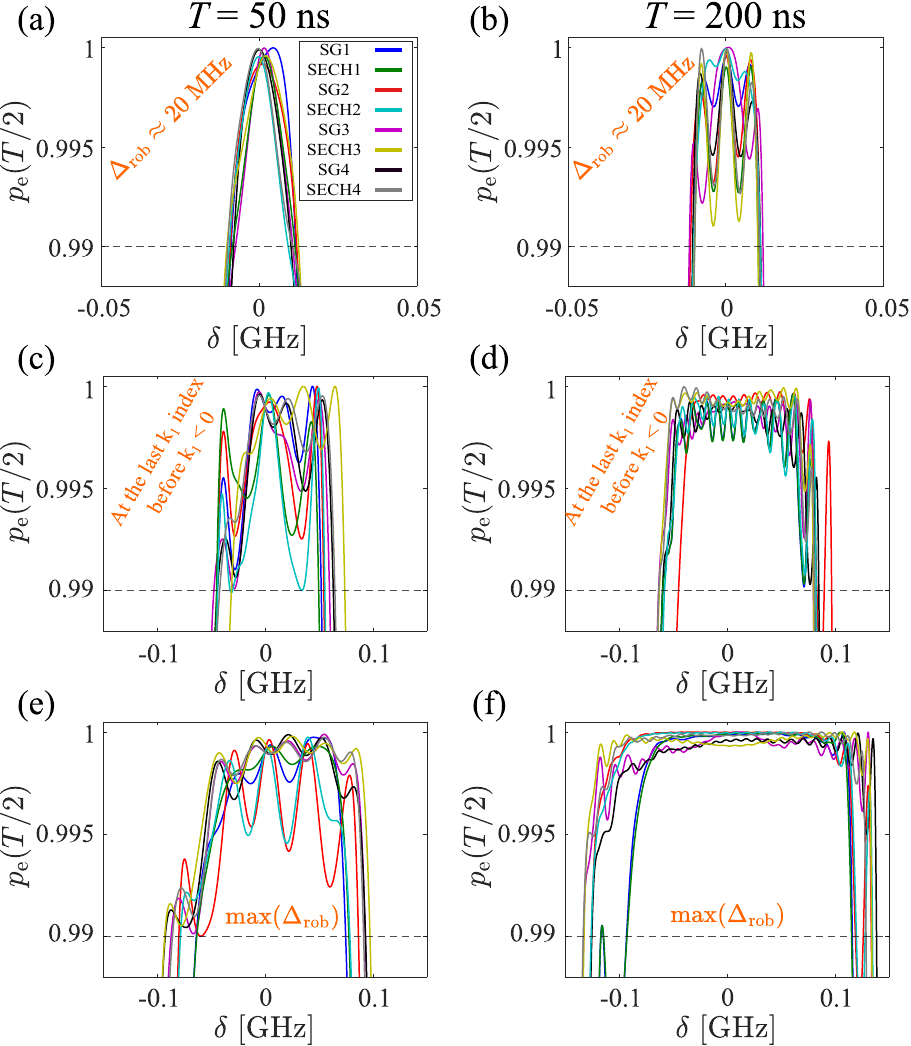}
\caption{Plots showing detuning sweeps for each of the models using the Pareto-optimal parameters. Panels (a) and (b) show detuning sweeps with a robustness of approximately $\Delta_{\rm rob} = 20$ MHz, for pulses of duration $T = 50$ ns and $T = 200$ ns, respectively. Panels (c) and (d) show the robustness of the models at the last $k_{1}$ index before $k_{1} < 0$, for $T = 50$ ns and $T = 200$ ns, respectively. Panels (e) and (f) display the sweeps with maximum detuning robustness for $T = 50$ ns and $T = 200$ ns, respectively. For reference, a black dashed line is plotted at $p_{\rm e}(T/2) = 0.99$ to indicate where the robustness (width) is measured.}%
\label{fig-det_sweeps}
\end{figure}

Fig.~\ref{fig-det_sweeps} shows the $p_{\rm e}(T/2)$ vs $\delta$ traces for different $k_1$ and pulse durations. For an adiabatic trajectory, the transfer probability is set only by the Landau-Zener velocity $v = \left. \frac{d \Delta (t)}{dt} \right|_{t=t_0}$ and the spectral gap $\propto \Omega(t_0)$ at the resonance condition $\Delta(t_0) + \delta = 0$. Since $\Delta(t)$ and $\Omega(t)$ are smooth functions of $t$, we expect $p_{\rm e}(T/2,\delta)$ to vary smoothly with respect to $\delta$. Instead, Fig.~\ref{fig-det_sweeps} reveals the presence of finer structure, which we attribute to non-adiabatic trajectories, where the full quantum dynamics set the outcome. Naturally, preserving adiabaticity is harder for a short pulse duration, leading to more fluctuations in the transfer probability as a function of $\delta$. This, in turn, decreases the total detuning robustness for short pulses. In contrast, the $T=200$ ns traces show little to no structure, implying an adiabatic population transfer. 

\FloatBarrier
\clearpage

\counterwithin{figure}{section}
\section{Amplitude Robustness}
\label{sec:Appendix_amprob}

Although amplitude robustness was not one of the optimization objectives, we used the solutions of the Pareto-optimization to compute it. Thus, the amplitude robustness naturally satisfies the same criterion used to define detuning robustness—namely, the width of the $p_{\rm{e}}(T/2)$ curve between the first two points where $p_{\rm{e}}(T/2) = 0.99$, provided that $p_{\rm{e}}(T/2) \geq 0.999$ is reached at least once within this interval.

\begin{figure}[h]
\centering
\includegraphics[width=0.5\linewidth]{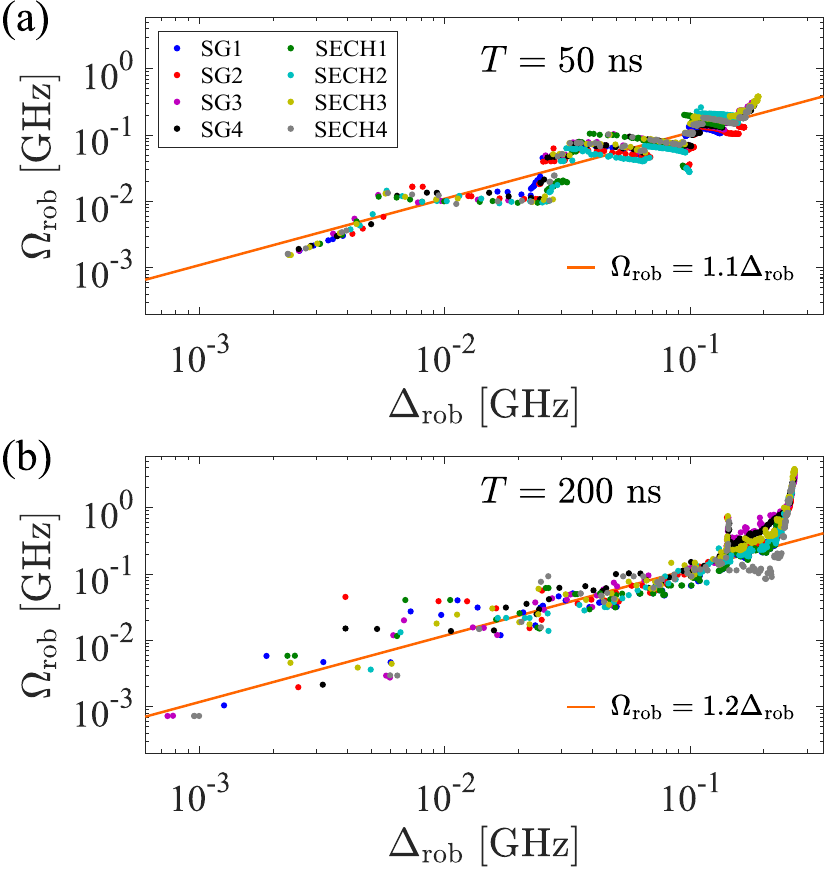}
\caption{A log-log scale plot showing amplitude robustness $\Omega_{\rm rob}$ vs detuning robustness $\Delta_{\rm rob}$ for each model evaluated at pulse durations (a) $T = 50$ ns and (b) $T = 200$ ns. The orange reference lines are the best-fit lines, each of which has a power of 1 and a slope near unity, indicating a near one-to-one relationship between $\Delta_{\rm rob}$ and $\Omega_{\rm rob}$.}%
\label{fig-amp_rob_vs_det_rob}
\end{figure}

Fig.~\ref{fig-amp_rob_vs_det_rob} shows a log-log scale of amplitude robustness versus detuning robustness for each model at the two pulse durations $T = 50$ ns (a) and $T = 200$ ns (b). The approximate linearity of the models and their convergence at higher robustnesses demonstrate that amplitude robustness and detuning robustness are highly correlated. Further, we see that the amplitude robustness $\Omega_{\rm rob}$ for the $T = 50$ ns pulses and the $T = 200$ ns pulses are each well-described by a linear relationship with respect to detuning robustness, i.e., $\Omega_{\rm rob} = 1.1\Delta_{\rm rob}$ and $\Omega_{\rm rob} = 1.2\Delta_{\rm rob}$, respectively.

\begin{figure}[h]
\centering
\includegraphics[width=0.65\linewidth]{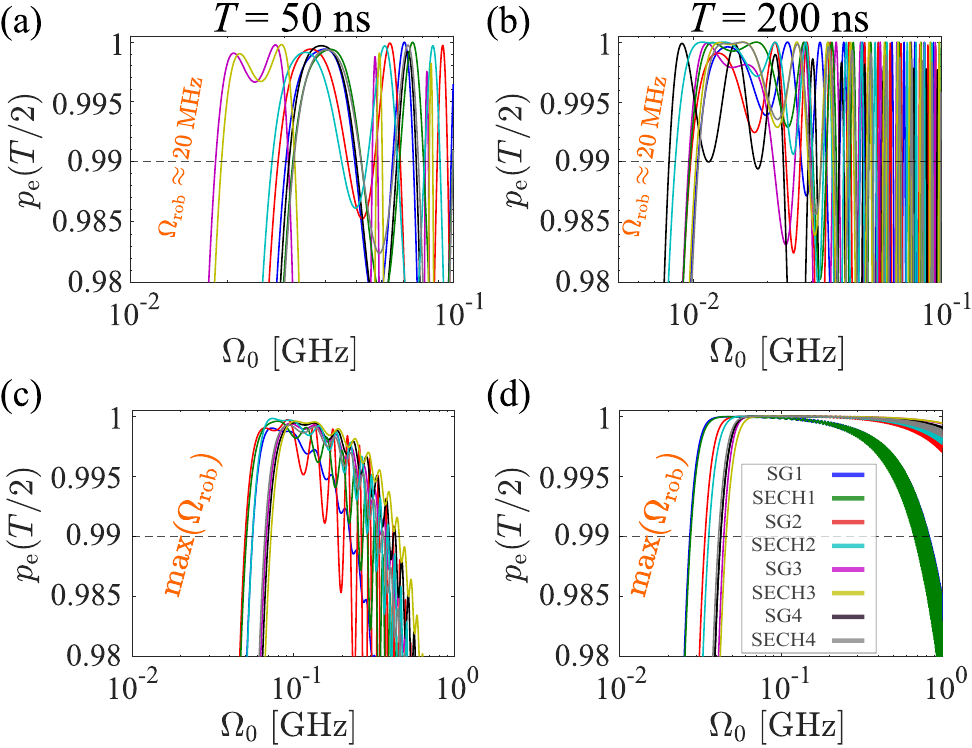}
\caption{Semi-log scale plots showing amplitude sweeps for each of the models using the Pareto-optimal parameters. Panels (a) and (b) show amplitude sweeps with a robustness of approximately $\Omega_{\rm rob} = 20$ MHz for pulses of duration $T = 50$ ns and $T = 200$ ns, respectively. Panels (c) and (d) display the sweeps with maximum amplitude robustness for $T = 50$ ns and $T = 200$ ns, respectively. For reference, a black dashed line is plotted at $p_{\rm e}(T/2) = 0.99$ to indicate where the robustness (width) is measured.}%
\label{fig-amp_sweeps}
\end{figure}

Finally, in Fig.~\ref{fig-amp_sweeps} we show the amplitude robustness for each model by plotting $p_{\rm e}(T/2)$ as a function of the amplitudes $\Omega_{0}$, for which we use a logarithmic scale to provide a better resolution. In particular, Fig.~\ref{fig-amp_sweeps}(a) and (b) show amplitude sweeps with effectively no robustness, i.e.,	 $\Omega_{\rm rob} \approx 20$ MHz for $T = 50$ ns and $T = 200$ ns pulses, respectively, and Fig.~\ref{fig-amp_sweeps}(c) and (d) show the traces with maximum amplitude robustness. Note that Fig.~\ref{fig-amp_sweeps}(d) which is for $T = 200$ ns pulses, only shows the full amplitude robustness for the models with linear detuning, as the nonlinear detuning models have greater amplitude robustness.

\FloatBarrier 

\end{appendices}

\bibliography{envelope_optimization.bib}

\end{document}